\documentclass[a4paper,10pt,twoside,fleqn,twocolumn]{article}
\usepackage{epsfig}
\usepackage{amssymb, amsmath}
\usepackage{graphicx}
\usepackage{subfigure}
\usepackage{url}
\usepackage[english]{babel}

\numberwithin{equation}{section}

\makeatletter
\newcommand\figcaption{\def\@captype{figure}\caption}
\newcommand\tabcaption{\def\@captype{table}\caption}
\makeatother

\renewenvironment{abstract}{\begin{quote}{\noindent\bf Abstract.}}{\end{quote}}

\begin{document}

  \title{
    \vspace*{-2cm}
       {\normalsize  Workload characterization and modelling
         \hfill \today \\[1mm]}
       \LARGE\sc\hrule height0.2pt \vskip 2mm
       Workload analysis of a cluster \\
       in a Grid environment
       \vskip 2mm \hrule height0.2pt}

  \author{ Emmanuel  Medernach\\[1ex]
    \small Laboratoire de Physique Corpusculaire, CNRS-IN2P3 \\
    \small Campus des C\'ezeaux, 
    \small 63177 Aubi\`ere Cedex, France \\
    \thanks{This work was supported by EGEE.} 
    \normalsize \em e-mail: \tt medernac@clermont.in2p3.fr }
  \date{}
  \maketitle

\begin{abstract}
  \em  With Grids, we  are able  to share  computing resources  and to
  provide for  scientific communities  a global transparent  access to
  local  facilities.  In  such  an environment  the  problems of  fair
  resource sharing and best usage  arise.  In this paper, the analysis
  of the LPC cluster usage (Clermont-Ferrand, France) in the EGEE Grid
  environment is done, and from the results a model for job arrival is
  proposed.
\end{abstract}

\section{Introduction}

 Analysis of a cluster workload  is essential to understand better user behavior
 and how  resources are  used~\cite{feitelson02workload}.  We are  interested to
 model  and simulate  the  usage of  a Grid  cluster  node in  order to  compare
 different scheduling policies and to find the best suited one for our needs.

 The Grid gives new ways to share resources between sites, both as computing and
 storage  resources.   Grid  defines   a  global  architecture  for  distributed
 scheduling  and  resource  management~\cite{darincosts} that  enable  resources
 scaling.  We would like to understand better  such a system so that a model can
 be defined.  With such a model, simulation may be done and a quality of service
 and fairness could then be proposed to the different users and groups.

 Briefly, we  have some  groups of  users that each  submit jobs  to a
 group of  clusters. These jobs are  placed inside a  waiting queue on
 some clusters before being  scheduled and then processed.  Each group
 of  users  have  their  own  need  and  their  own  strategy  to  job
 submittal. We wish:
 \begin{enumerate}
   \item to  have good metrics  that describes the  group and user usage  of the
     site.
   \item to model the global behavior (average job waiting time, average waiting
     queue  length, system  utilization, etc.)   in order  to know  what  is the
     influence of each parameter and to avoid site saturation.
   \item  to simulate  jobs arrivals  and  characteristics to  test and  compare
     different  scheduling   strategies.   The   goal  is  to   maximize  system
     utilization  and  to provide  fairness  between  site  users to  avoid  job
     starvation.
 \end{enumerate}

 As     parallel     scheduling    for     $p$     machines     is    a     hard
 problem~\cite{Garey79:intractability,      mertens04},      heuristics      are
 used~\cite{school-parallel,  feitelson95d}.  Moreover  we have  no  exact value
 about the duration of jobs, making the problem difficult.  We need a good model
 to be  able to compare different  scheduling strategies. We  believe that being
 able  to  characterize  users  and  groups  behavior  we  could  better  design
 scheduling  strategies that promote  fairness and  maintain a  good throughput.
 From this  paper some metrics are  revealed, from the job  submittal protocol a
 detailed arrival  model for  single user and  group is proposed  and scheduling
 problems are discussed.  We then suggest  a new design based on our observation
 and show relationship  between fairness issue and system  utilization as a flow
 problem.

 Our  cluster usage  in  the  EGEE (Enabling  Grids  for E-science  in
 Europe)  Grid  is presented  in  section~\ref{Environment}, the  Grid
 middleware  used is  described.  Corresponding  scheduling  scheme is
 shown  in section~\ref{Scheduling}.   Then  the workload  of the  LPC
 (Laboratoire  de  Physique   Corpusculaire)  computing  resource,  is
 presented (section~\ref{Workload analysis and modelling}) and the logs
 are   analyzed  statistically.    A   model  is   then  proposed   in
 section~\ref{Model}  that  describes the  job  arrival  rate to  this
 cluster.      Simulation    and     validation     are    done     in
 section~\ref{Simulation}  with  comparison   with  related  works  in
 section~\ref{related}.       Results       are      discussed      in
 section~\ref{Discussion}.   Section~\ref{Conclusion}  concludes  this
 paper.
   
 \section{Environment}
 \label{Environment}

 %% studied
 
 \subsection{Local situation}
 \label{Local situation}
 
 The EGEE node at  LPC Clermont-Ferrand is a Linux cluster made  of 140 dual 3.0
 GHz CPUs with 1 GB of RAM and  managed by 2 servers with the LCG (LHC Computing
 Grid  Project)  middleware.   We  are  currently  using  MAUI  as  our  cluster
 scheduler~\cite{jackson01, linux-usenix}.  It is  shared with the regional Grid
 INSTRUIRE (\url{http://www.instruire.org}).  Our LPC Cluster role in EGEE is to
 be used mostly by Biomedical users\footnote{Our cluster represented 75\% of all
 the Biomed Virtual  Organization (VO) jobs in 2004.}  located  in Europe and by
 High Energy  Physics Communities.  Biomedical research is  one core application
 of the EGEE project.  The approach  is to apply the computing methods and tools
 developed in  high energy  physics for biomedical  applications.  Our  team has
 been involved  in international research group focused  on deploying biomedical
 applications in a Grid environment.

 %% Biomedical  applications currently  used in the  Biomed Virtual Organization  (VO) are
 %% listed on table~\ref{table:applications}. 

 One   pilot  application  is   GATE  which   is  based   on  the   Monte  Carlo
 GEANT4~\cite{geant4} toolkit  developed by  the high energy  physics community.
 Radiotherapy and brachytherapy use ionizing radiations to treat cancer.  Before
 each treatment,  physicians and physicists plan the  treatment using analytical
 treatment planning systems  and medical images data of the  tumor. By using the
 Grid environment  provided by the EGEE project,  we will be able  to reduce the
 computing time of Monte Carlo simulations in order to provide a reasonable time
 consuming tool for specific cancer treatment requiring Monte-Carlo accuracy.

 Another group is  Dteam, this group is partly  responsible of sending
 tests and monitoring  jobs to our site.  Total CPU  time used by this
 group is  small relatively to  the other one,  but the jobs  sent are
 important for the  site monitoring.  There are also  groups using the
 cluster from the  LHC experiments at CERN (\url{http://www.cern.ch}).
 There are  different kind  of jobs for  a given group.   For example,
 Data Analysis  requires a lot  of I/O whereas  Monte-Carlo Simulation
 needs few I/O.
 
 \subsection{EGEE Grid technology}
 \label{Grid technology}

 In Grid world, resources are controlled by their owners. For instance
 different kind of scheduling policies could be used for each site.  A
 Grid resource  center provides to  the Grid computing  and/or storage
 resources and also services that allow jobs to be submitted by guests
 users,  security  services, monitoring  tools,  storage facility  and
 software management.  The main issue  of submitting a job to a remote
 site   is  to  provide   some  warranty   of  security   and  correct
 execution. In  fact the  middleware automatically resubmits  job when
 there is  a problem with  one site.  Security and  authentication are
 also provided as Grid services.

 The Grid principle is to allow user a worldwide transparent access to
 computing and storage resources.  In the case of EGEE, this access is
 aimed to be  transparent by using LCG middleware built  on top of the
 Globus Toolkit~\cite{foster97globus}.  Middleware  acts as a layer of
 software that provides homogeneous  access to different Grid resource
 centers.

 \subsection{LCG Middleware}
 \label{LCG Middleware}

 LCG   is  organized   into  Virtual   Organizations   (VOs):  dynamic
 collections of  individuals and  institutions sharing resources  in a
 flexible,  secure  and   coordinated  manner.   Resource  sharing  is
 facilitated and controlled by a  set of services that allow resources
 to be  discovered, accessed, allocated, monitored  and accounted for,
 regardless of their physical location. Since these services provide a
 layer  between physical  resources and  applications, they  are often
 referred to as Grid Middleware~\cite{glite-arch}.

 Bag  of  task  applications  are parallel  applications  composed  of
 independent  jobs.  No  communications are  required  between running
 jobs.  Since  jobs from  a same task  may execute on  different sites
 communications  between jobs  are  avoided.  In  this context,  users
 submit their jobs to the Grid one by one through the middleware.  Our
 cluster receives jobs  only from the Grid.  This  means that each job
 requests  for  one and  only  one  processor.   Users could  directly
 specify  the execution site  or let  a Grid  service choose  the best
 destination  for them.   Users give  only a  rough estimation  of the
 maximum  job  running  time.   In  general  this  estimated  time  is
 overestimated  and very  imprecise~\cite{zotkin99joblength}.  Instead
 of  speaking about an  estimated time,  it could  be better  to speak
 about  an upper bound  for job  duration, so  this value  provided by
 users is  more a precision value.   The bigger the value  is the more
 imprecise the value of the actual runtime could be.

\begin{figure*}[ht]
   \centering
   \includegraphics[height=12.3cm]{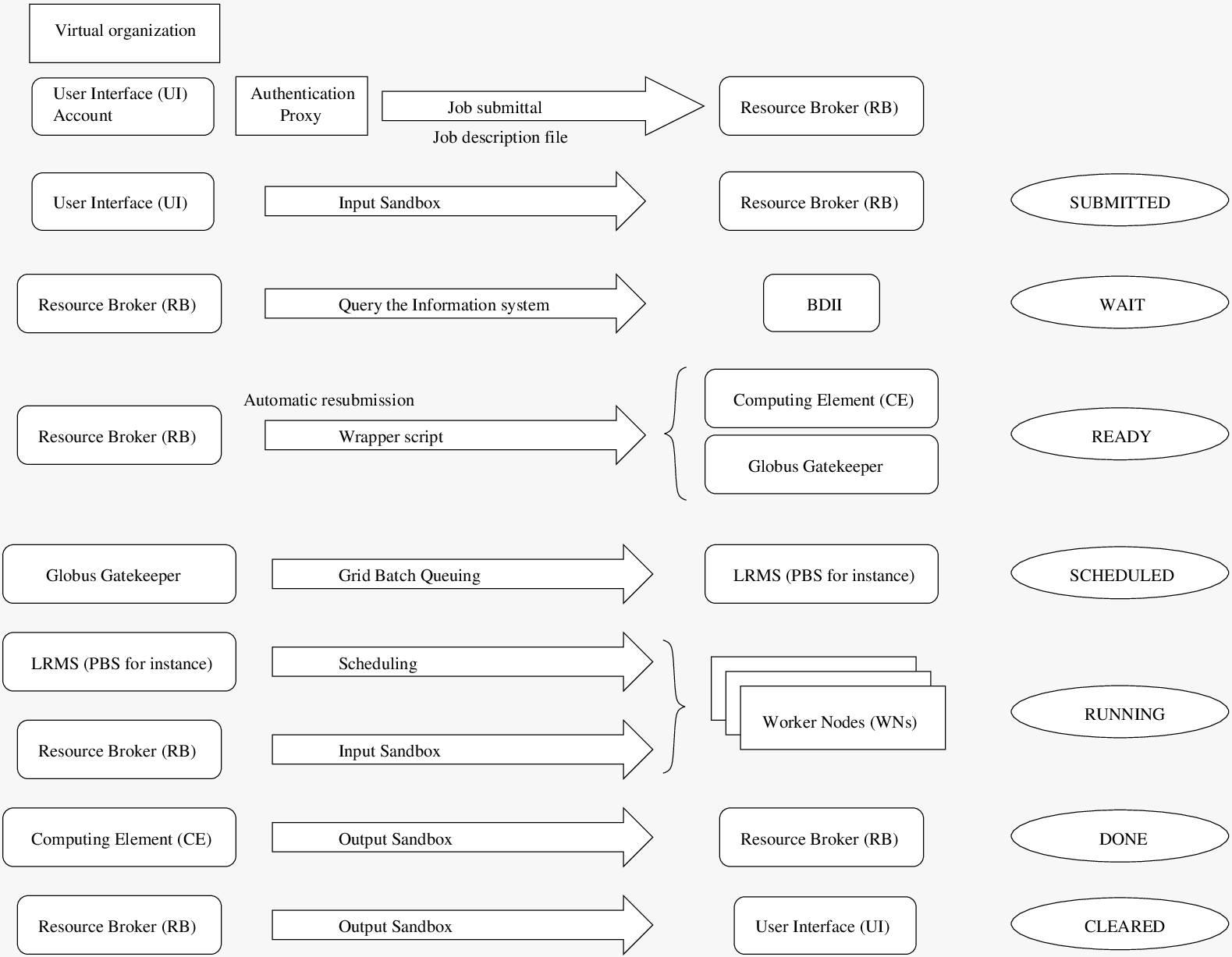}
   \caption{Job submittal scenario}
   \label{fig:jobflow}
 \end{figure*}

 Figure~\ref{fig:jobflow} shows  the scenario  of a job  submittal. In
 this  figure rounded  boxes are  grid services  and ellipses  are the
 different jobs  states. As there  is no communications  between jobs,
 jobs  could  run  independently  on multiple  clusters.   Instead  of
 communicating  between  job execution,  jobs  write  output files  to
 Storage Elements (SE) of the  Grid.  Small output files could also be
 sent to  the UI.   Replica Location Service  (RLS) is a  Grid service
 that allow location of replicated data.  Other jobs may read and work
 on the data generated, forming ``pipelines'' of jobs.
 
 The  users Grid entry  point is  called an  User Interface  (UI).  This  is the
 gateway to Grid services.  From this machine, users are given the capability to
 submit   jobs   to   a   Computing   Element   and   to   follow   their   jobs
 status~\cite{lcguserguide}. A Computing Element  (CE) is composed of Grid batch
 queues.  A Computing Element is built  on a homogeneous farm of computing nodes
 called Worker Nodes (WN) and on a node called a GateKeeper acting as a security
 front-end to the rest of the Grid.

 %% Ok

 Users can query the Information System in order  to know both the state of different grid nodes and
 where their jobs are able to run depending on job requirements.  This match-making process has been
 packaged as a Grid service known as the  Resource Broker (RB). Users could either submit their jobs
 directly to  different sites or to  a central Resource Broker  which then dispatches  their jobs to
 matching sites.

 The services  of the Workload Management  System (WMS) are  responsible for the
 acceptance of job submits and the  dispatching of these jobs to the appropriate
 CEs, depending  on job requirements  and on available resources.   The Resource
 Broker is the machine where the WMS  services run, there is at least one RB for
 each  VO.  The  duty  of the  RB  is to  find the  best  resource matching  the
 requirements   of   a   job   (match-making  process).    (For   more   details
 see~\cite{dataGrid-wms})

 Users are then mapped to a  local account on the chosen executing CE.
 When a CE receives a job,  it enqueues it inside an appropriate batch
 queue,  chosen  depending  on  the  job  requirements,  for  instance
 depending on the maximum running time.  A scheduler then proceeds all
 these queues to  decide the execution of jobs.   Users could question
 about status of their jobs during all the job lifetime.

 \section{Scheduling scheme}
 \label{Scheduling}

 The goal  of the scheduler is  first to enable execution  of jobs, to
 maximize job  throughput and to  maintain a good  equilibrium between
 users in their usage of the cluster~\cite{feitelson96c}.  At the same
 time scheduler has to avoid starvation, that is jobs, users or groups
 that  access  scarcely to  available  cluster  resources compared  to
 others.

 Scheduling is done on-line, i.e  the scheduler has no knowledge about
 all the job  input requests but jobs are submitted  to the cluster at
 arbitrary  time.    No  preemption  is  done,  the   cluster  uses  a
 space-sharing mode for jobs.  In a Grid environment long-time running
 jobs are common.  The worst case  is when the cluster is full of jobs
 running for days  and at the same time receiving  jobs blocked in the
 waiting queue.

 Short jobs  like monitoring jobs  barely delay too much  longer jobs.
 For example, a  1 day job could wait 15  minutes before starting, but
 it is  unwise if a  5 minutes  job has to  wait the same  15 minutes.
 This results  in production of algorithms classes  that encourage the
 start  of short  jobs  over  longer jobs.   (Short  jobs have  higher
 priority~\cite{chiang02})  Some other solution  proposed is  to split
 the cluster in static sub-clusters  but this is not compatible with a
 sharing  vision like  Grids.  Ideal  on-line scheduler  will maximize
 cluster usage  and fairness  between groups and  users.  Of  course a
 good trade-off has to be found between the two.

 \subsection{Local situation}
 
 We are  using two  servers to  manage our 140  CPUs, on  each machine
 there are 5  queues where each group could send  their jobs to.  Each
 queue has its own limit in maximum  CPU Time.  A job in a given queue
 is killed if it exceeds its  queue time limit.  There are in fact two
 limits, one  is the maximum  CPU time, the  other one is  the maximum
 total time (or  Wall time) a job could use.  For  each queue there is
 also a  limit in the  number of  jobs than can  run at a  given time.
 This is  done in order  to avoid that  the cluster is full  with long
 running jobs and short jobs  cannot run before days.  Likely there is
 the same limit in number of running jobs for a given group.
 
 \begin{table}[h]
   \begin{center}
   \begin{tabular}{|l|c|c|c|}
     \hline
     Queue & Max CPU & Max Wall & Max Jobs \\
           & (H:M) & (H:M) & \\
     \hline
     Test        & 00:05 & 00:15 & 130 \\
     Short       & 00:20 & 01:30 & 130 \\
     Long        & 08:00 & 24:00 & 130 \\
     Day         & 24:00 & 36:00 & 130 \\
     Infinite    & 48:00 & 72:00 & 130 \\
     \hline
   \end{tabular}
   \end{center}
   \caption{Queue  configuration  (maximum  CPU  time, Wall  time  and
   running jobs)}
 \end{table}

Maui Scheduler  and the  Portable Batch System  (PBS) run  on multiple
hardware and  operating systems.  MAUI  is a scheduling  policy engine
that is used together with the  PBS batch system.  PBS manages the job
reception  in  queues and  execution  on  cluster  nodes.  MAUI  is  a
First-Come-First-Served  backfill  scheduler  with  priorities.   This
means  that is checks  periodically the  running queues,  execution of
lower priority jobs is allowed  if it is determined that their running
will not delay jobs  higher in the queue~\cite{linux-usenix}.  Maui is
unfortunately not event  driven, it polls regularly the  PBS queues to
decide which jobs to run.  MAUI  allows to add a priority property for
each  queue.  Our  site configuration  is that  the shorter  the queue
allows jobs to run, the more  priority is given to that job.  Jobs are
then  selected  to  run depending  on  a  priority  based on  the  job
attributes such as owner, group, queue, waiting time, etc.

%% If a job violates a site policy it is placed temporary in a blocked
%% state and not considered for scheduling. 

 \section{Workload data analysis}
 \label{Workload analysis and modelling}

 %% - Ce qu'on observe -> modele qui approche le comportement
 
Workload    analysis    allows    to    obtain    a   model    of    the    user
behavior~\cite{calzarossa93workload}.    Such   a   model   is   essential   for
understanding  how the different  parameters change  the resource  center usage.
Meta-computing workload~\cite{chapin99b}  like Grid environments  is composed of
different site workloads.   We are interested in modelling  workload of our site
which  is part  of the  EGEE computational  Grid.  Our  site receives  only jobs
coming from the EGEE Grid and each requests for only one CPU.

Traces of users activities are obtained from accountings on the server logs.  Logs contain
information about users,  resources used, jobs arrival time and  jobs completion time.  It
is possible to use directly these traces to obtain a static simulation or to use a dynamic
model  instead.  Workload  models  are more  flexible  than logs,  because  they allow  to
generate   traces    with   different   parameters   and    better   understand   workload
properties~\cite{feitelson02workload}.   Workload analysis  allows  to obtain  a model  of
users activity.  Such a model is  essential for understanding how the different parameters
change the  resource center usage.  Our workload  data has been converted  to the Standard
Workload Format (\url{http://www.cs.huji.ac.il/labs/parallel/workload/}) and made publicly
available for further researches.

\begin{figure*}[hp]
  \begin{center}
    \subfigure[Number of Biomed jobs received per weeks (from August 2004 to May 2005)]{\label{fig:stats:jobWeekBiomed}\includegraphics[totalheight=5.2cm,clip]{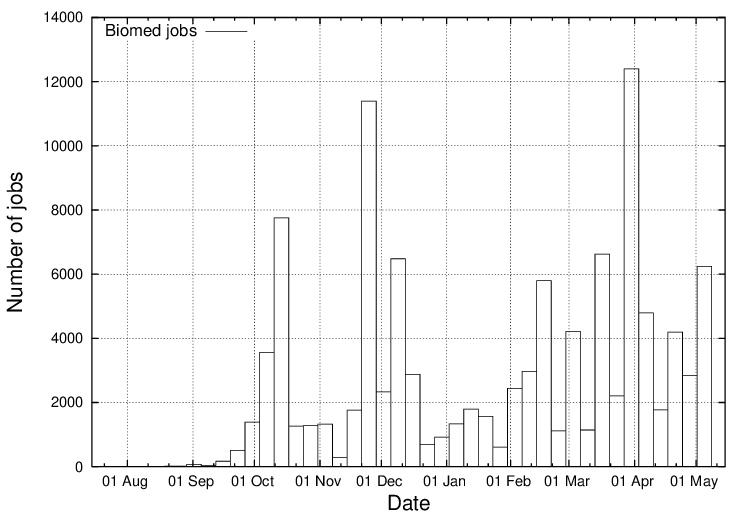}} 
    \subfigure[Number of Dteam jobs received per weeks (from August 2004 to May 2005)]{\label{fig:stats:jobWeekDteam}\includegraphics[totalheight=5.2cm,clip]{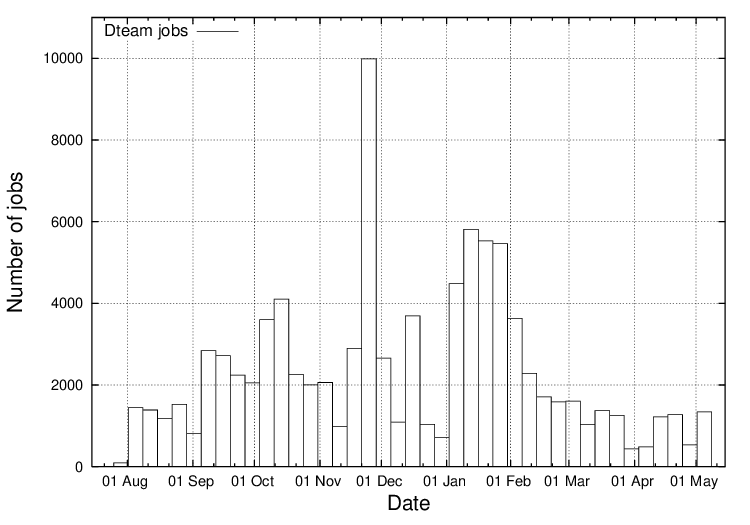}}
    \caption{Number of jobs  received per VO and per  week from August
    2004 to May 2005}
    \label{fig:stats}
  \end{center}
\end{figure*}
\begin{figure*}[hp]
  \begin{center}
    \subfigure[System utilization per weeks (from August 2004 to May 2005)]{\label{fig:stats:utilization}\includegraphics[totalheight=5.2cm,clip]{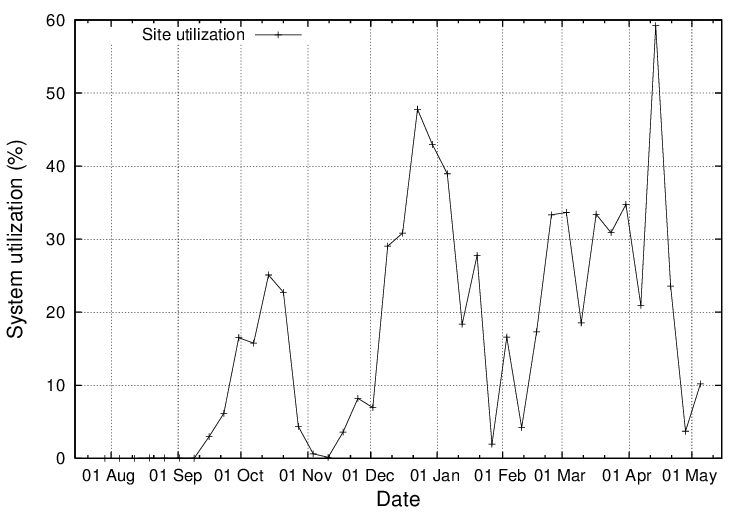}}  
    \subfigure[CPU consumed by Biomed and Dteam jobs per weeks (from August 2004 to May 2005)]{\label{fig:stats:cpuBiomed}\includegraphics[totalheight=5.2cm,clip]{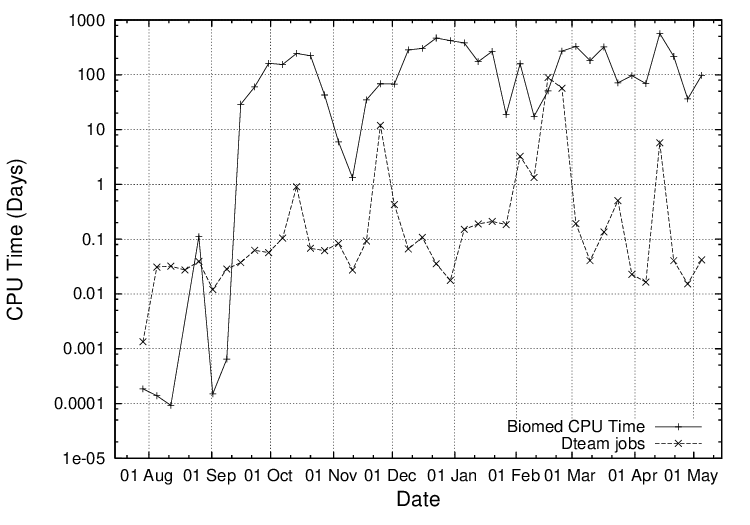}} \\
%%    \subfigure[CPU consumed by Dteam jobs per weeks (from August 2004 to May 2005)]{\label{fig:stats:cpuDteam}\includegraphics[totalheight=4.6cm,clip]{NumberDteamCPUPerWeek.eps}} 
%%    \subfigure[CPU consumed by Atlas jobs per weeks (from August 2004 to May 2005)]{\label{fig:stats:cpuAtlas}\includegraphics[totalheight=4.6cm,clip]{NumberAtlasCPUPerWeek.eps}} \\
%%    \subfigure[CPU consumed by LHCb jobs per weeks (from August 2004 to May 2005)]{\label{fig:stats:cpuLHCb}\includegraphics[totalheight=4.6cm,clip]{NumberLHCbCPUPerWeek.eps}} \\
    \caption{Cluster utilization as CPU consumed per VO and per week from August 2004 to May 2005}
    \label{fig:statscpu}
  \end{center}
\end{figure*}

Workload is from August 1st 2004  to May 15th 2005.  We have a cluster
containing 140 CPUs since September  15th.  This can be visible in the
figure~\ref{fig:stats},         \ref{fig:stats:utilization}        and
\ref{fig:stats:cpuBiomed},  where we  notice that  the number  of jobs
sent increases.  Statistics are obtained  from the PBS log files.  PBS
log files  are well  structured for data  analysis.  An AWK  script is
used to  extract information  from PBS log  files.  AWK acts  on lines
matched  by regular  expressions.  We  do not  have  information about
users \emph{Login} time because users send jobs to our cluster from an
User Interface (UI) of the EGEE Grid and not directly.

\subsection{Running time}

\begin{figure*}[ht]
  \begin{center}
    \subfigure[Dteam job runtime]{\label{figure:dteam:jobduration}\includegraphics[totalheight=5.2cm,clip]{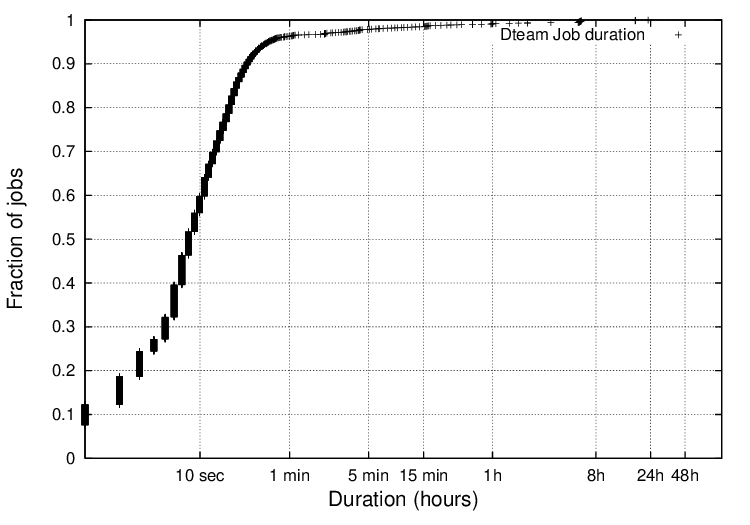}} 
    \subfigure[Biomed job runtime]{\label{figure:biomed:jobduration}\includegraphics[totalheight=5.2cm,clip]{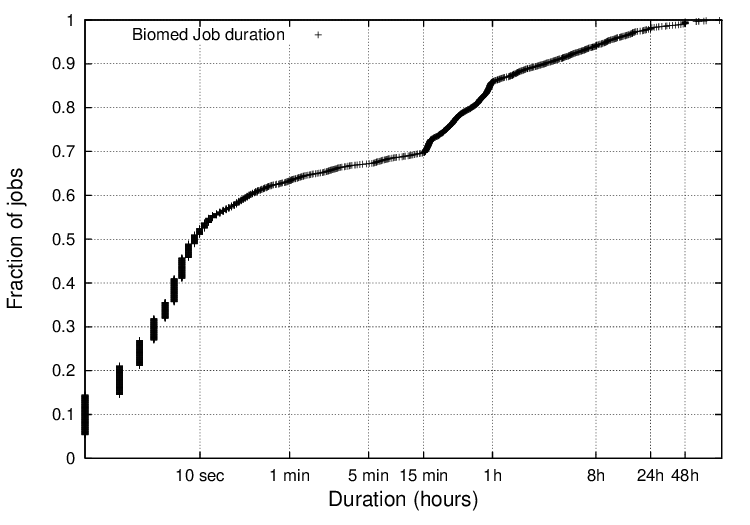}} 
    \caption{Dteam and Biomed job runtime distributions (logscale on time axis)}
    \label{fig:group:runtime}
  \end{center}
\end{figure*}

\begin{table}[h]
  \begin{center}
    \begin{tabular}{|l|c|c|c|}
      \hline
      Group & Mean & Standard & Number \\
      &  & Deviation & of jobs \\
      \hline
      Biomed & 5417 & 22942.2 & 108197 \\
      Dteam & 222 & 3673.6 & 94474 \\
      LHCb & 2072 & 7783.4 & 9709 \\
      Atlas & 13071 & 28788.8 & 7979 \\
      Dzero & 213 & 393.9 & 1332 \\
      \hline
    \end{tabular}
  \end{center}
  \caption{Group  running time  in seconds  and total  number  of jobs
  submitted}
  \label{table:group:runtime}
\end{table}

During 280 days,  our site received 230474 jobs  from which 94474 Dteam jobs and  108197 Biomed jobs
(table~\ref{table:group:runtime}).  For  all these jobs  there are 23208  jobs that failed  and were
dequeued.   It appears  that jobs  are submitted  irregularly and  by bursts,  that is  lot  of jobs
submitted in  a short period of  time followed by a  period of relative inactivity.   From the logs,
there are not  much differences between CPU time and  total time, so it means that  jobs sent to our
cluster are really CPU intensive jobs and not I/O intensive.  Dteam jobs are mainly short monitoring
jobs but  all Dteam jobs  are not  regularly sent jobs.   We have 6784.6  days CPU time  consumed by
Biomed  for 108197  jobs (Mean  of  one hour  and half  per jobs,  table~\ref{table:group:runtime}).
Repartition   of   cumulative   job   duration   distributions   for   Biomed   VO   is   shown   on
figure~\ref{fig:group:runtime}.  The duration of about 70\%  of Biomed jobs are less than 15 minutes
and 50\% under 10 seconds, there are a dominant number of small running jobs but the distribution is
very   wide   as   shown   by   the    high   standard   deviation   compared   to   the   mean   in
table~\ref{table:group:runtime}.

\begin{table}[h]
  \begin{center}
    \begin{tabular}{|l|c|c|c|}
      \hline
      Queue &  Mean & Standard & CV \\
      & & Deviation & \\
      \hline
      Test  & 31.0 & 373.6 & 12.0 \\
      Short  & 149.5 & 1230.5 & 8.2 \\
      Long  & 2943.2 & 11881.2 & 4.0 \\
      Day  & 6634.8 & 25489.2 & 3.8 \\
      Infinite  & 10062.2 & 30824.5 & 3.0 \\
      \hline
    \end{tabular}
  \end{center}
  \caption{Queue  mean  running  time  in  seconds,  corresponding
  Standard Deviation and Coefficient of Variation}
  \label{table:queue:runtime}
\end{table}

Users  submit   their  jobs  with   an  estimated  run   length.   For
relationships between  execution time  and requested job  duration and
its  accuracy see~\cite{cirnecompr-model}.  To  sum up  estimated jobs
duration are essentially inaccurate. It  is in fact an upper bound for
job duration which could in reality take any value below it.
Table~\ref{table:queue:runtime} shows for  each queue the mean running
time, its  standard deviation and coefficient of  variation (CV) which
is the ratio between standard deviation and the mean.  CV decreases as
the queue maximum  runtime increase.  This means that  jobs in shorter
queues vary a lot in their duration compared to longer jobs and we can
expect that more the upper bound  given is high the more confidence in
using the queue mean runtime as a an estimation we could have.

A commonly used  method for modelling duration distribution  is to use
log-uniform    distribution.    Figures~\ref{figure:dteam:jobduration}
and~\ref{figure:biomed:jobduration}  show the  fraction  of Dteam  and
Biomed jobs with duration less than some value.  Job duration has been
modelled      with      a      multi-stage      log-uniform      model
in~\cite{downey99elusive} which is piecewise  linear in log space.  In
this  case  Dteam  and  Biomed  job  duration  could  be  approximated
respectively with a 3 and a 6 stages log-uniform distribution.

\subsection{Waiting time}

\begin{table}[h]
  \begin{center}
    \begin{tabular}{|l|c|c|c|c|}
      \hline
      Group & Mean & Stretch & Standard & CV \\
      &  & & Deviation &  \\
      \hline
      Biomed & 781.5   & 0.874 & 16398.8 & 20.9  \\
      Dteam  & 1424.1  & 0.135 & 26104.5 & 18.3  \\
      LHCb   & 217.7   & 0.905 & 2000.7 & 9.1  \\
      Atlas  & 2332.8  & 0.848 & 13052.1 & 5.5  \\
      Dzero  & 90.7    & 0.701 & 546.3 & 6.0   \\
      \hline
    \end{tabular}
  \end{center}
  \caption{Group mean waiting  time in seconds, corresponding Standard
    Deviation and Coefficient of Variation}
  \label{table:waiting}
\end{table}
 
\begin{table}[h]
  \begin{center}
    \begin{tabular}{|l|c|c|c|c|}
      \hline
      Queue & Mean & Standard & CV & Number \\
      &  & Deviation &  & of jobs \\
      \hline
      Test & 33335.9 & 148326.4 & 4.4 & 45760 \\
      Short & 1249.7 & 27621.8 & 22.1 & 81963 \\
      Long & 535.1 & 5338.8 & 9.9 & 32879 \\
      Day & 466.8 & 8170.7 & 17.5 & 19275 \\
      Infinite & 1753.9 & 24439.8 & 13.9 & 49060 \\
      \hline
    \end{tabular}
  \end{center}
  \caption{Queue mean waiting  time in seconds, corresponding Standard
    Deviation, Coefficient of Variation and number of jobs}
  \label{fig:queuemean}
\end{table}

Table~\ref{table:waiting} shows that jobs  coming from the Dteam group
are  the more  unfairly treaten.   Dteam group  sends short  jobs very
often, Dteam jobs are then all  placed in queue waiting that long jobs
from other groups finished.  Dzero  group sends short jobs more rarely
and is  also less  penalized than Dteam  because there are  less Dzero
jobs that  are waiting  together in queue  before being  treated.  The
best treated group is LHCb with not very long running jobs (average of
about  34 minutes)  and  one job  about  every 41  minutes.  The  best
behavior to reduce  waiting time per jobs seems to  send jobs that are
not too  short compared to the  waiting factor, and send  not too very
often in  order to avoid that  they all wait together  inside a queue.
Very long jobs is not a  good behavior too as the scheduler delay them
to run shorter one if possible.

Table~\ref{fig:queuemean} shows  the mean waiting  time per jobs  on a
given  queue.  There is  a problem  with such  a metric,  for example:
Consider one job arriving on a cluster with only one free CPU, it will
run on it  during a time $T$ with no waiting  time.  Consider now that
this job  is splitted  in $N$ shorter  jobs (numbered $0  \ldots N-1$)
with  equal total duration  $T$.  Then  the waiting  time for  the job
number $i$ will be $iT/N$,  and the total waiting time $(N-1)T/2$.  So
the more  a job is splitted the  more it will wait  in total.  Another
metric that does not depend on the number of jobs is the total waiting
time divided by the number of jobs and by the total job duration.  Let
note $\widehat{WT}$ this normalized waiting time, We obtain:
\begin{align}
\widehat{WT} & = \frac{TotalWaitingTime}{NJobs * TotalDuration} \nonumber\\
\widehat{WT} & = \frac{MeanWaitingTime}{NJobs * MeanDuration} 
\end{align}

 \begin{table}[h]
   \begin{center}
   \begin{tabular}{|l|c||l|c|}
     \hline
     Queue & $\widehat{WT}$ & Group & $\widehat{WT}$ \\
     \hline
     Test & 2.35e-2 & Biomed & 1.58e-5 \\
     Short & 1.02e-4 & Dteam & 6.79e-5 \\
     Long & 5.53e-6 & LHCb & 1.08e-5 \\
     Day & 3.65e-6 & Atlas & 2.23e-5 \\
     Infinite & 3.55e-6 & Dzero & 31.9e-5 \\
     \hline
   \end{tabular}
   \end{center}
   \caption{Queue and Group normalized waiting time}
 \end{table}
 With this  metric, the Test  queue is  still the most  unfairly treated and  the Infinite
 queue has  the more  benefits compared  to the other  queues.  Dteam  group is  again bad
 treated because their jobs are mainly sent  to the Test queue.  The more unfairly treated
 group is Dzero.

%% Todo
%% In fact, sending short jobs  rarely is the worst behavior for general
%% waiting time as it is unlikely  that two jobs will wait together in a
%% queue.
 
 \subsection{Arrival time}

 \begin{figure}[h]
   \centering \includegraphics[totalheight=5.2cm]{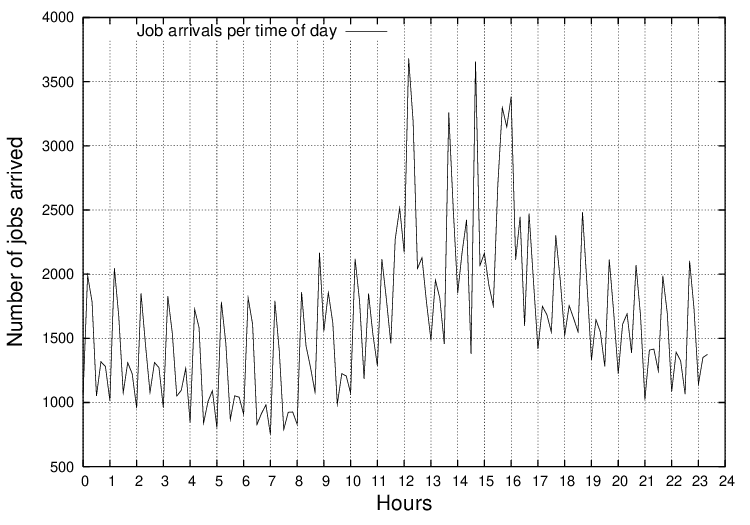}
   \caption{Job arrival daily cycle}
   \label{fig:arrivalhours}
 \end{figure}
 
 \begin{table}[h]
   \begin{center}
     \begin{tabular}{|l|c|c|c|}
       \hline
       Group & Mean & Standard & CV\\
       & (seconds) & Deviation & \\
       \hline
       Biomed & 223.6 & 5194.5 & 23.22 \\
       Dteam & 256.2 & 2385.4 & 9.31 \\
       LHCb & 2474.6 & 39460.5 & 15.94 \\
       Atlas & 2824.1 & 60789.4 & 21.52 \\
       Dzero & 5018.7 & 50996.6 & 10.16 \\
       \hline
     \end{tabular}
   \end{center}
   \caption{Group  interarrival  time  in  seconds,  corresponding
     Standard Deviation and Coefficient of Variation}
   \label{table:interarrival}
 \end{table}

Job arrival daily cycle is presented in figure~\ref{fig:arrivalhours}.
This  figure shows  the number  of  arrival depending  on job  arrival
hours, with a 10 minutes sampling.  Clearly users prefer to send their
jobs at o'clock.  In fact  we receive regular monitoring jobs from the
VO  Dteam.   The  monitoring   jobs  are  submitted  every  hour  from
\url{goc.grid-support.ac.uk}.  Users are located in all Europe, so the
effect of sending at working hours is summed over all users timezones.
However the  shape is  similar compared to  other daily  cycle, during
night (before  8am) less jobs are  submitted and there  is an activity
peak around midday, 2pm and 4pm.

Table~\ref{table:interarrival} shows the  moments of interarrival time
for each group.  CV is much  higher than $1$, this means that arrivals
are not  Poisson processes and are very  irregularly distributed.  For
instance  we could  receive  $10$  jobs in  $10$  minutes followed  by
nothing during  the $50$ next  minutes.  In this  case we have  a mean
interarrival time  of $6$ minutes but  in fact when  jobs arrived they
arrived every minutes.

Figure~\ref{fig:stats:utilization} shows the system utilization of our
cluster  during  each week.   There  are a  maximum  of  980 CPU  days
consumed  each week for  140 CPUs.  We have  a highly  varying cluster
activity.

\subsection{Frequency analysis}

\begin{figure*}[ht]
  \begin{center}
    \subfigure[Biomed job arrival rate each 5 minutes]{\label{table:stats:rateBiomed}\includegraphics[totalheight=5.2cm,clip]{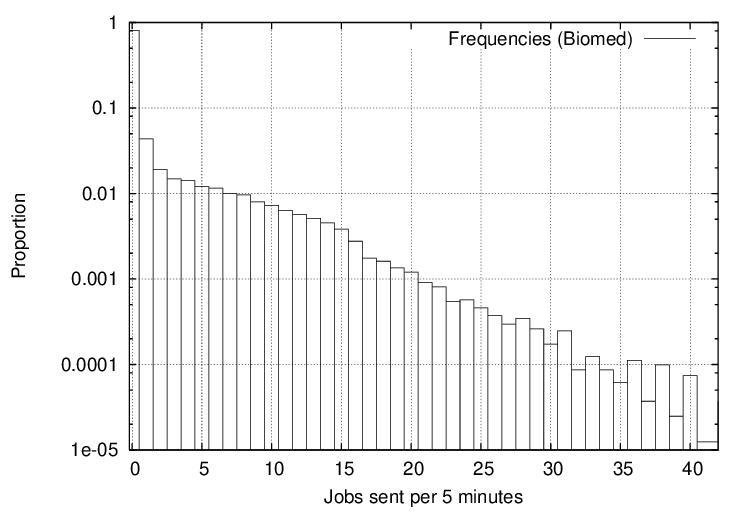}} 
    \subfigure[Dteam job arrival rate each 5 minutes]{\label{table:stats:rateDteam}\includegraphics[totalheight=5.2cm,clip]{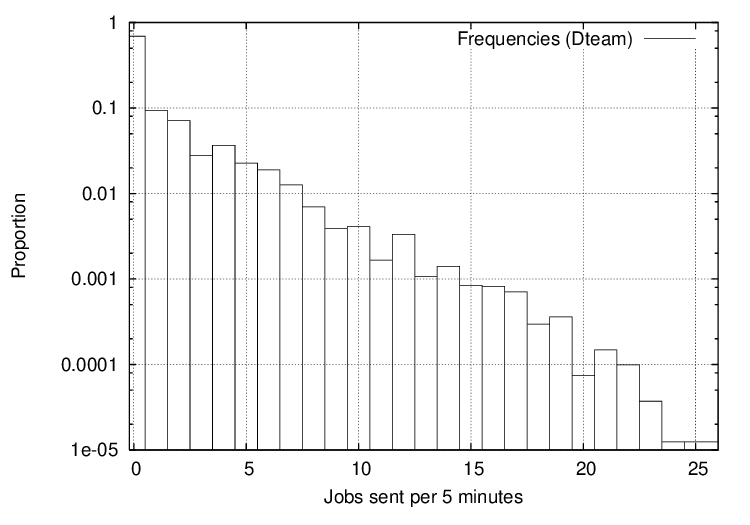}}   
    \caption{Arrival frequencies for  Biomed and Dteam VOs (Proportion
      of  occurrences of  $n$ jobs  received during  an interval  of 5
      minutes)}
  \label{fig:stats:more}
  \end{center}
\end{figure*}

Job  arrival  rate  is  a  common   measurement  for  a  site  usage  in  queuing  theory.
Figures~\ref{table:stats:rateBiomed}   and~\ref{table:stats:rateDteam}  present   the  job
arrival  rate distribution.   It  is the  number of  time  $n$ jobs  are submitted  during
interval  length of 5  minutes.  They  show that  most of  the time  the cluster  does not
receive  jobs but jobs  arrived grouped.   Users actually  submit groups  of jobs  and not
stand-alone jobs.  It explains the shape of  the arrival rate: it fastly decreases but too
slowly  compared to  a Poisson  distribution.  Poisson  distribution is  usually  used for
modelling the arrival process but evidences are against that fact~\cite{feitelson95e}.

 Dteam monitoring jobs are short and regular jobs, there is no need of a special
 arrival model for  such jobs.  What we  observe for other kind of  jobs is that
 the job  arrival law is  not a Poisson Law  (see table~\ref{table:interarrival}
 where $CV \gg 1$) as for instance a web site traffic~\cite{paxson95wide}.  What
 really happens  is that  users come  using the cluster  from an  User Interface
 during some  time interval.  During  this time they  send jobs to  the cluster.
 Users log to an User Interface machine in order to send their jobs to a RB that
 dispatch them  to some CEs.  Note  that one can  send jobs to our  cluster only
 from an  User Interface, it means for  instance that jobs running  on a cluster
 cannot send secondary jobs.  On a computing site we do not have this user login
 information, but only job arrival.
 
 First  we look  at modelling  user arrival  and  submission behavior.
 Secondly we  show that  the model proposed  shows good results  for a
 group behavior.

 \section{Model}
 \label{Model}
 
 \subsection{Login model}
 \label{Login model}

 In this section we begin to model user \emph{Login}/\emph{Logout} behavior from
 the Grid  job flow  (figure~\ref{fig:jobflow}).  We neglect  the case  where an
 user has multiple login on different UI  at the same time.  We mean that a user
 is in the state  \emph{Login} if he is in the state of  sending jobs from an UI
 to our cluster, else he is in the state \emph{Logout}.

\begin{figure}[h]
  \centering
  \includegraphics[height=5.2cm]{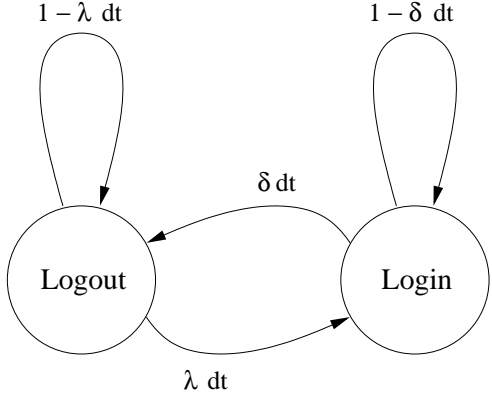}
  \caption{\emph{Login/Logout} cycle}
  \label{fig:login}
\end{figure}

Markov  chains  are  like  automatons  with  for each  state  a  probability  of
transition.  One property of Markov chains  is that future states depend only on
the current state and not on the  past history.  So a Markov state must contains
all  the  information  needed  for  future  states.  We  decided  to  model  the
\emph{Login/Logout} behavior as a continuous  Markov chain.  During each $dt$, a
\emph{Logout} user has a probability during  $dt$ of $\lambda dt$ to login and a
\emph{Login} user  has a probability during  $dt$ of $\delta dt$  to logout (see
figure~\ref{fig:login}).  $\lambda$ is called the \emph{Login} rate and $\delta$
is called the \emph{Logout} rate.

All these parameters could vary over time as we see with the variation
of the week job arrival (figure~\ref{fig:stats:utilization}) or during
day time  (figure~\ref{fig:arrivalhours}) The model  proposed could be
used more  accurately with non-constant  parameters at the  expense of
more calculation  and more difficult fitting.  For  example, one could
numerically use  Fourier series for  the \emph{Login} rate or  for the
submittal  rate  to model  this  daily  cycle.   We use  now  constant
parameters for calculation, looking for general properties.

We would like  to have the probabilities during time  that the user is
logged    or   not    logged.    Let    $\mathcal{P}_{Login}(t)$   and
$\mathcal{P}_{Logout}(t)$ be respectively probability that the user is
logged or not logged at time t. We have from the modelling:
\begin{eqnarray}
  \lefteqn{\mathcal{P}_{Logout}(t+dt) = (1 - \lambda dt) \mathcal{P}_{Logout}(t)}\nonumber  \\
  & & \mbox{} + \delta dt \mathcal{P}_{Login}(t) \\
  \lefteqn{\mathcal{P}_{Login}(t+dt) = (1 - \delta dt) \mathcal{P}_{Login}(t)}\nonumber \\
  & & \mbox{} + \lambda dt \mathcal{P}_{Logout}(t) 
\end{eqnarray}
At equilibrium we have no variation so
\begin{eqnarray}
  \mathcal{P}_{Logout}(t+dt) = \mathcal{P}_{Logout}(t) = \mathcal{P}_{Logout} \\
  \mathcal{P}_{Login}(t+dt) = \mathcal{P}_{Login}(t) = \mathcal{P}_{Login}
\end{eqnarray}
We obtain: 
\begin{align}
  \mathcal{P}_{Logout} &= \frac{\delta}{\lambda + \delta} \\
  \mathcal{P}_{Login} &= \frac{\lambda}{\lambda + \delta}
\end{align}

\subsection{Job submittal model}
\label{Job submittal model}

During period  when users are logged they  could submit jobs.  We  model the job
submittal rate for one  user as follows: During $dt$ when the  user is logged he
has  a probability  of $\mu  dt$ to  submit a  job.  With  $\delta=0$ we  have a
delayed Poisson process, with $\mu=0$ no  jobs are submitted.  The full model is
shown  at  figure~\ref{fig:markov}, it  shows  all  the  possible outcomes  with
corresponding probabilities from  one of the possible state to  the next after a
small period $dt$.  Numbers inside circles are the number of jobs submitted from
the start.   \emph{Login} states are below  and \emph{Logout} states  are at the
top.  We have:

\begin{figure*}[ht]
  \centering
  \includegraphics[height=8.6cm]{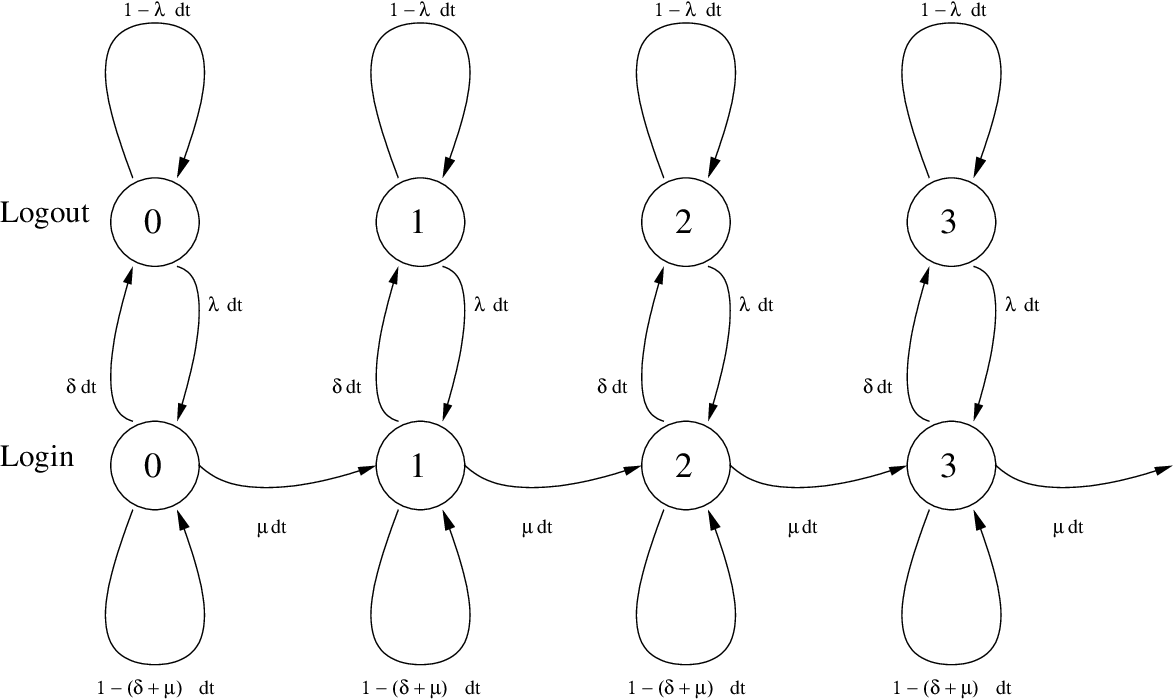}
  \caption{Markov modelling of jobs submittal}
  \label{fig:markov}
\end{figure*}

\begin{itemize}
\item  $\mathcal{P}_n(t)$  is  the  probability  to be  in  the  state
  ``\emph{User is not logged at time  t and n jobs have been submitted
  between time 0 and t}.''
\item  $\mathcal{Q}_n(t)$  is  the  probability  to be  in  the  state
  ``\emph{User  is logged at  time t  and n  jobs have  been submitted
  between time 0 and t}.''
\item  $\mathcal{R}_n(t)$  is  the  probability  to be  in  the  state
  ``\emph{n jobs have been submitted  between time 0 and t}.'' We have
  $\mathcal{R}_n = \mathcal{P}_n + \mathcal{Q}_n$.
\end{itemize}

From  the  model,  we obtain  with  the  same  method as  before  this
recursive differential equation:
\begin{eqnarray}
  \mathcal{M} & = &
    \begin{pmatrix}
      - \lambda & \delta \\
      \lambda & -(\mu + \delta)
    \end{pmatrix} \\  
  \begin{pmatrix}
    \mathcal{P}_0 \\
    \mathcal{Q}_0
  \end{pmatrix} ' & = &
  \mathcal{M}
  \begin{pmatrix}
    \mathcal{P}_0 \\
    \mathcal{Q}_0
  \end{pmatrix} \\
  \begin{pmatrix}
    \mathcal{P}_n \\
    \mathcal{Q}_n
  \end{pmatrix} ' & = &
  \mathcal{M}
  \begin{pmatrix}
    \mathcal{P}_n \\
    \mathcal{Q}_n
  \end{pmatrix} +
  \begin{pmatrix}
    0 \\
    \mu  \mathcal{Q}_{n-1}
  \end{pmatrix}
\label{recursive}
\end{eqnarray}

This  results  to  the  following  recursive  equation  (in  case  the
parameters are constants, $\mathcal{M}$ is a constant)
\begin{eqnarray}
  \begin{pmatrix}
    \mathcal{P}_n \\
    \mathcal{Q}_n
  \end{pmatrix} & = &
  e^{\mathcal{M}t} 
  \int{
    e^{-\mathcal{M}x} 
    \begin{pmatrix}
      0 \\
      \mu  \mathcal{Q}_{n-1}
    \end{pmatrix}
    dx}
\end{eqnarray}
    
We take a look at the probability of having no job arrival during an interval of
time $t$  which is $\mathcal{P}_0$ and $\mathcal{Q}_0$.   $\mathcal{R}_0$ is the
the probability that no jobs have been submitted between arbitrary time 0 and t.
So from the above model, we have:
\begin{equation}\label{eqmatrix2}
  \begin{pmatrix}
    \mathcal{P}_0 \\
    \mathcal{Q}_0
  \end{pmatrix} ' =
  \begin{pmatrix}
    - \lambda & \delta \\
    \lambda & -(\mu + \delta)
  \end{pmatrix} 
  \begin{pmatrix}
    \mathcal{P}_0 \\
    \mathcal{Q}_0
  \end{pmatrix} 
\end{equation}

At  arbitrary  time  we  could  be  in  the  state  \emph{Login}  with
probability  $\lambda   /  (\lambda  +  \delta)$  and   in  the  state
\emph{Logout} with the probability  $\delta / (\lambda + \delta)$.  We
have from the results above: 
\begin{align}
  \begin{pmatrix}
    \mathcal{P}_0(0) \\
    \mathcal{Q}_0(0)
  \end{pmatrix} &=
  \begin{pmatrix}
    \mathcal{P}_{Logout} \\
    \mathcal{P}_{Login}
  \end{pmatrix} =
  \frac{1}{\lambda + \delta}
  \begin{pmatrix}
    \delta \\
    \lambda
  \end{pmatrix}
  \\
  \mathcal{R}_0 &= \mathcal{P}_0 + \mathcal{Q}_0
\end{align}

Finally we obtain the result. 
\begin{equation}
\begin{split}
  \mathcal{R}_0(t) &= m_0 \frac{e^{- m_1t} - e^{- m_2t}}{m_1
    - m_2} + \frac{m_1e^{- m_2t} - m_2e^{- m_1t}}{m_1 - m_2} 
\end{split}
\end{equation}

Where \begin{align}
  m_0 &= \frac{\lambda \mu}{\lambda + \delta} = \mu \mathcal{P}_{Login} \\
  m_1 + m_2 &= \lambda + \delta + \mu \\
  m_1 m_2 &= \lambda \mu \\
\end{align}

With $\lambda=0$ or $\mu=0$, we  obtain that no jobs are submitted $(\mathcal{R}_0(t)=1)$.
With $\delta=0$, this  is a Poisson process and  $\mathcal{R}_0(t)=e^{-\mu t}$.  Note that
during a period of $t$ there are in average $\mu \mathcal{P}_{Login} t$ jobs submitted, we
have also for small period $t$,
\begin{equation}
  \mathcal{R}_0(t) \approx 1 - \mu\mathcal{P}_{Login}t
\end{equation} 
We have also
\begin{align}
\mathcal{R}'_0(0) & = - \mu\mathcal{P}_{Login} \\
\mathcal{R}'_0(0) & = - \frac{\emph{Number of jobs submitted}}{\emph{Total duration}}
\end{align}

$\mathcal{R}_0(t)$  could be  estimated by  splitting the  arrival  processes in
intervals  of  duration  $t$ and  estimating  the  ratio  of intervals  with  no
arrival. The error of this estimation  is linear with $t$. Another issue is that
the logs precision is not below one second. 

\subsection{Model characteristics}

We have also these interesting properties:
\begin{eqnarray}
  \mathcal{R}'_0(0) & = & - \mu \mathcal{P}_{Login} \\
  \frac{\mathcal{R}''_0(0)}{\mathcal{R}'_0(0)} & = & - \mu \\
  \frac{\mathcal{R}'_0(0)^{2}}{\mathcal{R}''_0(0)} & = & \mathcal{P}_{Login} \\
  \frac{\mathcal{R}'''_0(0)}{\mathcal{R}''_0(0)} & = & - (\mu + \delta)
\end{eqnarray}

Probability  distribution  of  the  duration   between  two  jobs  arrival  is  called  an
interarrival process.  Interarrival process is a common metric in queuing theory.  We have
$\mathcal{A}(t)=\mathcal{P}_0(t)+\mathcal{Q}_0(t)$  with the  initial condition  that user
just submits a job.  This implies that user is logged.
\begin{equation}
  \mathcal{P}_0(0) = 0.0, \quad \mathcal{Q}_0(0) = 1.0 \nonumber
\end{equation}
\begin{equation}
\begin{split}
\mathcal{A}(t)  &= \mu \frac{e^{- m_1t} -  e^{- m_2t}}{m_1 -
    m_2} + \frac{m_1e^{- m_2t} - m_2e^{- m_1t}}{m_1 - m_2}
\end{split}
\end{equation}
\begin{equation}
  p = \frac{\mu - m_2}{m_1 - m_2} 
\end{equation}
\begin{equation}
  \mathcal{A}(t) = p e^{- m_1t} + (1 - p) e^{- m_2t} 
\end{equation}
We have $\mu \in[m1; m2]$ because
\begin{align}
  (\mu - m_1)(\mu - m_2) & = \mu^{2} - (\lambda + \delta + \mu)\mu + \delta \mu \nonumber\\
  (\mu - m_1)(\mu - m_2) & = -\delta \mu < 0
\end{align}
So $p \in  [0; 1]$, and we have  an hyper-exponential interarrival law
of  order 2 with  parameters $p=(\mu  - m_2)/(m_1  - m_2),  m_1, m_2$.
This   result   is    coherent   with   other   experimental   fitting
results~\cite{david-workload}  Moreover any  hyper-exponential  law of
order  2  may   be  modelled  with  the  Markov   chain  described  in
figure~\ref{fig:markov}  with parameters  $\mu =  p m_1  +  (1-p) m_2,
\lambda = m_1 m_2 / \mu, \delta = m_1 + m_2 - \mu - \lambda$

Let calculate the  mean interarrival time.  Probability to  have an interarrival
time   between   $\theta$  and   $\theta+d\theta$   is  $\mathcal{A}(\theta)   -
\mathcal{A}(\theta+d\theta) = -\mathcal{A}'(\theta)d\theta$.  The mean is
\begin{eqnarray}
\tilde{\mathcal{A}} &=& \int_{0}^{\infty}-\theta \mathcal{A}'(\theta)d\theta = \int_{0}^{\infty}\mathcal{A}(\theta)d\theta \\
\tilde{\mathcal{A}} &=& \frac{1}{\mu \mathcal{P}_{Login}} = \frac{\lambda+\delta}{\lambda\mu} 
\end{eqnarray}

Let compute the variance of interarrival distribution. 
\begin{eqnarray}
  \mathit{var} &=& \int_{0}^{\infty}-(\theta - \tilde{\mathcal{A}})^{2}\mathcal{A}'(\theta)d\theta \\
  \mathit{var} &=& 2 \int_{0}^{\infty}\theta\mathcal{A}(\theta)d\theta - \tilde{\mathcal{A}}^{2} \\
  \frac{\mathit{var}}{\tilde{\mathcal{A}}^{2}} &=& {CV}^{2} = 1 + 2\frac{\delta\mu}{(\lambda+\delta)^{2}} \\
  {CV}^{2} &=& 1 + 2~\mathcal{P}_{Logout}^{2} \frac{\mu}{\delta} 
\end{eqnarray}

Another  interesting property  is the  number of  jobs submitted  by this  model  during a
\emph{Login}  period.   Let  $P_n$  be  the  probability to  receive  $n$  jobs  during  a
\emph{Login} period. We have:
\begin{eqnarray}
  P_n & = & \int_{0}^{\infty} \frac{(\mu t)^{n}}{n!}e^{-\mu t} \delta e^{-\delta t} dt \\
  P_n & = & \frac{\delta}{\mu + \delta}(\frac{\mu}{\mu + \delta})^{n}
\end{eqnarray}
This  is  a  geometric law.  The  mean  number  of jobs  submitted  by
\emph{Login} period is $\mu/\delta$.

\subsection{Group model}

Groups are composed of users, either  regular users sending jobs at regular time
or users with a \emph{Login/Logout} like behavior.  Metrics defined below as the
mean number  of jobs  sent by  \emph{Login} state, the  mean submittal  rate and
probability of \emph{Login} could represent an user behavior.

\begin{figure}[h]
  \centering
  \includegraphics[height=5.2cm]{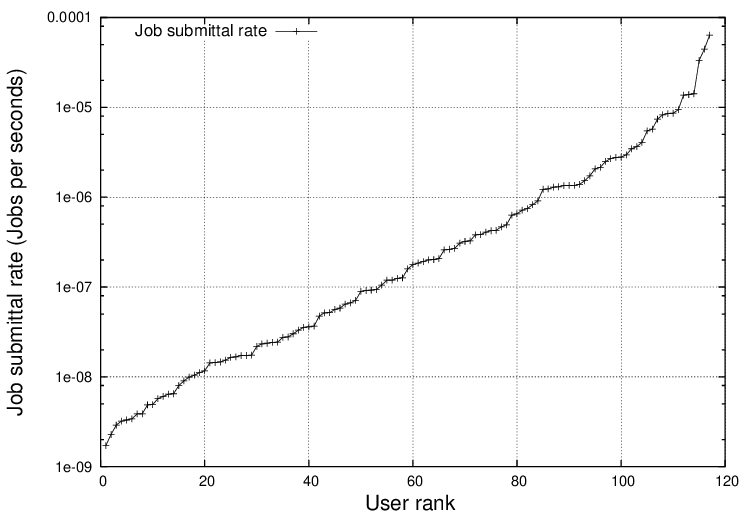}
  \caption{Users job submittal rates during their period of activity}
  \label{fig:jobrate}
\end{figure}

Figure~\ref{fig:jobrate} shows  the sorted distribution of  users submittal rate
($\mu  \mathcal{P}_{Login}$).  Except  for  the  highest values  it  is quite  a
straight line in logspace.  This observation could be included in a group model.

\section{Simulation and validation}
\label{Simulation}

 %% - Evaluation des parametres: r0(0) connu avec precision = µPlogin = nbjobs/tempstotal
 %% - µPlogin classé par utilisateur, variance: conjecture d'une loi de groupe
 %% - Graphe des R0 de plusieurs utilisateurs
 %% - Norme utilisée, Fitting des fréquences

\begin{figure*}[hp]
  \begin{center}
   \subfigure[Biomed user 1]{\label{figure:simu:biomed001.clrce02}\includegraphics[totalheight=5.4cm,clip]{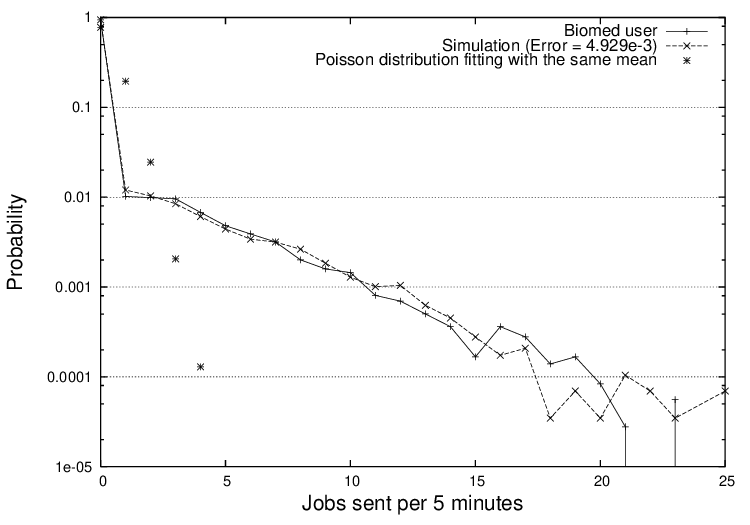}}
   \subfigure[Biomed user 2]{\label{figure:simu:biomed001.clrce01}\includegraphics[totalheight=5.4cm,clip]{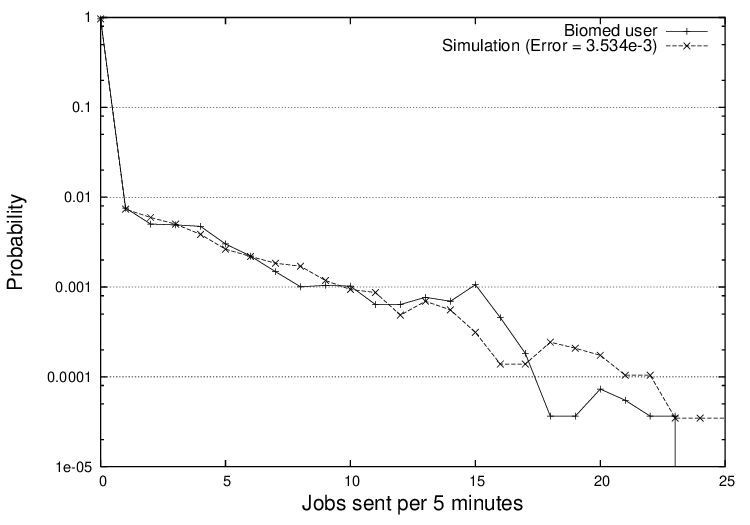}} \\
   \subfigure[Biomed user 3]{\label{figure:simu:biomed002.clrce01}\includegraphics[totalheight=5.4cm,clip]{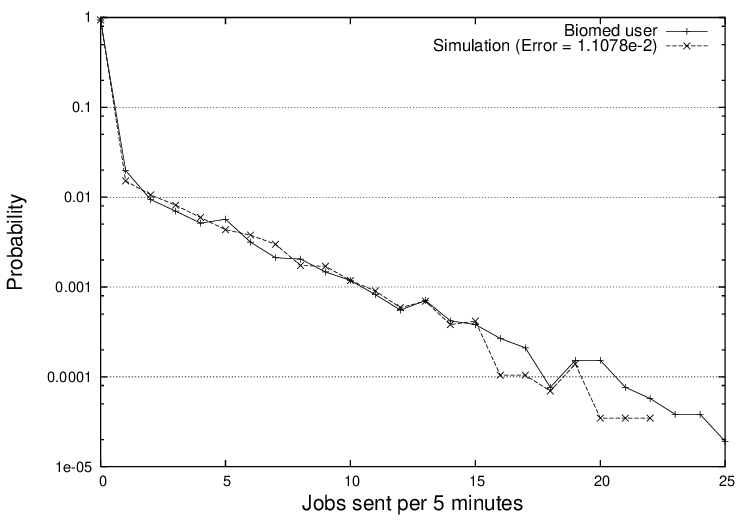}} 
   \subfigure[Biomed user 4]{\label{figure:simu:biomed005.clrce01}\includegraphics[totalheight=5.4cm,clip]{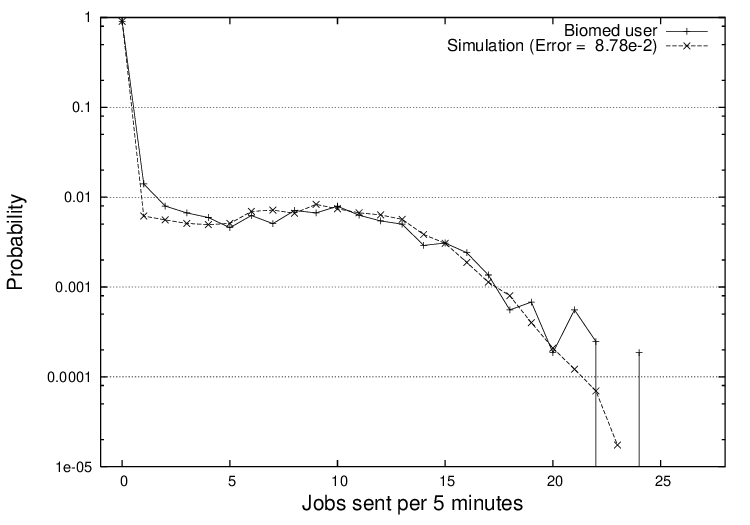}}    \\
    \begin{tabular}{|l | c | c | c | c |}
      \hline
      Name & $\mu$ & $\delta$ & $\lambda$ & Error \\ 
      \hline
      Biomed user 1 & 0.0837 & 0.02079 & 2.1e-4 & 4.929e-3 \\
      Biomed user 2 & 0.0620 & 0.01188 & 1.2e-4 & 3.534e-3 \\
      Biomed user 3 & 0.0832 & 0.02475 & 2.5e-4 & 1.1078e-2 \\
      Biomed user 4 & 0.0365 & 1.4285e-3 & 1.075e-4 & 8.78e-2\\
      \hline
    \end{tabular}
  \end{center}
  \caption{Biomed simulation results}
  \label{table:simu}
\end{figure*}

\begin{figure*}[hp]
  \begin{center}
    \subfigure[LPC cluster Biomed user]{\label{fig:r0Biomed}\includegraphics[totalheight=5.4cm,clip]{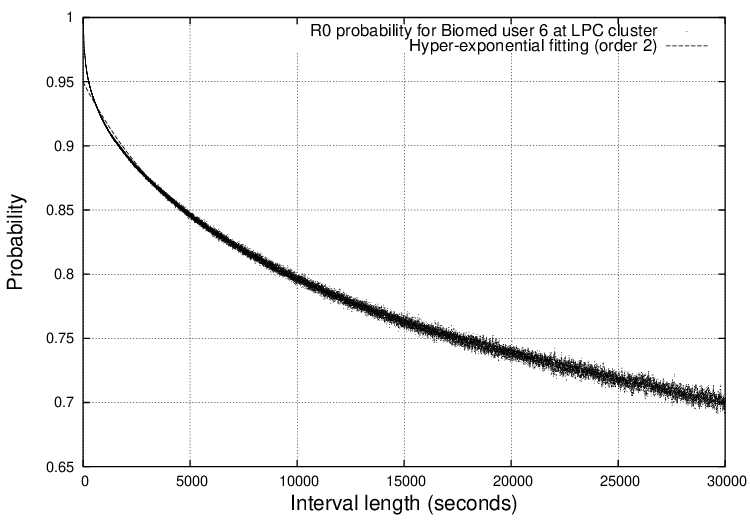}}
    \subfigure[NASA Ames most active user]{\label{fig:nasa:nasa}\includegraphics[totalheight=5.4cm,clip]{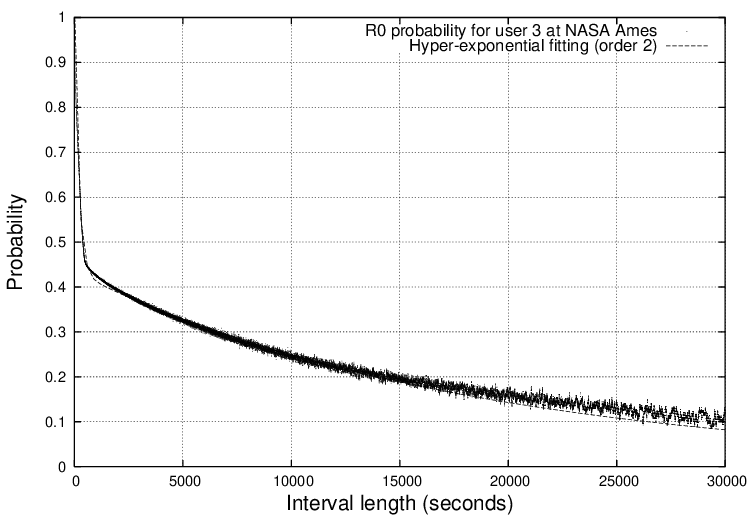}} \\
    \subfigure[DAS2 fs0 cluster most active user]{\label{fig:nasa:das2}\includegraphics[totalheight=5.4cm,clip]{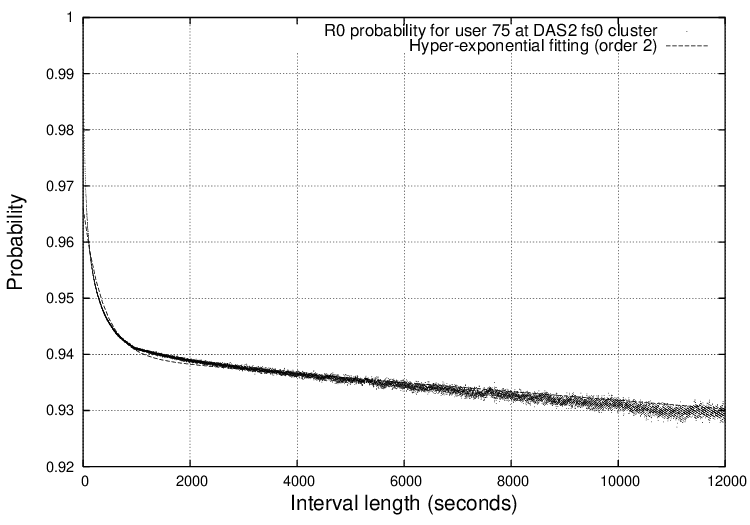}}
    \subfigure[SDSC Blue Horizon most active user]{\label{fig:nasa:sdsc}\includegraphics[totalheight=5.4cm,clip]{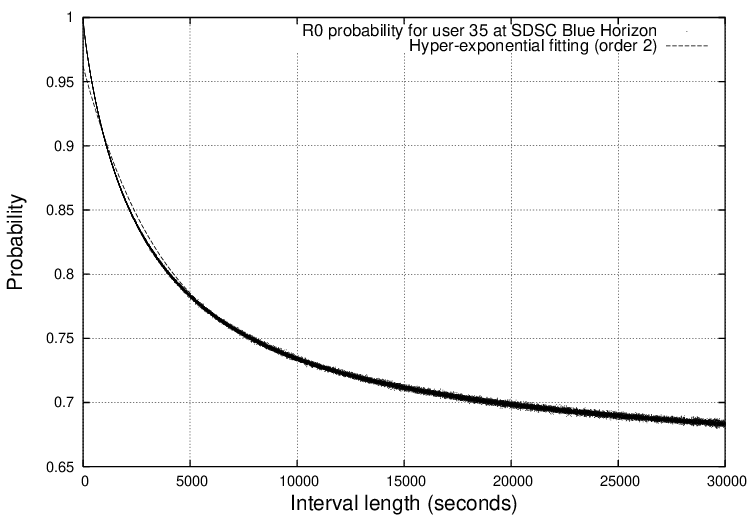}}
    \caption{Hyper-Exponential fitting of $\mathcal{R}_0$ for a Biomed
      LPC user  and for the most  active users at NASA  Ames, DAS2 and
      SDSC Blue Horizon clusters.}
    \label{fig:nasa}
  \end{center}
\end{figure*}

We have  done a simulation  in Scheme~\cite{kelsey98revised} directly  using the
Markov model.  We began by fitting  users behavior from the logs with our model.
Like  the frequency  obtained  from the  logs,  the model  shows  a majority  of
intervals with no job arrival, possibly followed by a relatively flat area and a
fast     decreases.     Some     fitting     results    are     presented     in
figures~\ref{table:simu}. Norm used  to fit real data is  the maximum difference
between the two cumulative distributions.  We fitted the frequency data for each
user.

During a  period of $t$  there are in  average $\mu \mathcal{P}_{Login}  t$ jobs
submitted.   We evaluate  the value  of $\mu  \mathcal{P}_{Login}$ which  is the
average number of jobs submitted by seconds.  We use that value when doing a set
of simulation  in order to fit  a known real user  probability distribution.  We
have two free  parameters, so we vary $\mathcal{P}_{Login}$  between 0.0 and 1.0
and   $lambda$   which  the   inverse   is  the   average   time   an  user   is
\emph{Logout}. Some results obtained are shown in figures~\ref{table:simu}.

$\mu$ parameter decides of the frequency length of the curve.  Without
the  \emph{Login} behavior we  would have  obtained a  classic Poisson
curve of $\mu$ parameter. $1/\mu$ is the mean interarrival time during
\emph{Login} period.  An  idea to evaluate $\mu$ would  be to evaluate
the job  arrival rate  during \emph{Login} periods,  but we  lack that
\emph{Login} information.

$\delta$ and $\lambda$ are the  \emph{Logout} and \emph{Login} parameters.  What is really
important is the ratio $\lambda/(\lambda  + \delta)$ which is $\mathcal{P}_{Login}$.  This
is the  ratio between time  user is active  on the cluster  and total time.   $\delta$ and
$\mathcal{P}_{Login}$  are measures  of the  deviation from  a classic  Poisson  law.  For
instance, the mean number of jobs submitted by \emph{Login} period is $\mu/\delta$ and the
mean job submittal rate is  $\mu\mathcal{P}_{Login}$.  For a same $\mathcal{P}_{Login}$ we
could have  very different scenarios.   A user  could be active  for long time  but rarely
logged and another user could be  active for short period with frequent login.  $1/\delta$
is the mean \emph{Login} time, $1/\lambda$ is the mean \emph{Logout} time.

The  $\mathcal{R}_0$  probability  is essential  for  studying  job  arrival time.   $1  -
\mathcal{R}_0(t)$ is  the probability that  between time $0$  and $t$ we have  received at
least one job.  It is easier to  fit the $\mathcal{R}_0$ distribution for an user than the
interarrival distribution because we  have more points.  Figure~\ref{fig:r0Biomed} shows a
typical  graph of  $\mathcal{R}_0$ for  a Biomed  user.  It  shows for  instance  that for
intervals of 10000 seconds, this Biomed user has a probability of about 0.2 to submits one
or more  jobs.  We have  fitted this probability  with hyper-exponential curve, that  is a
summation of exponential curves.  There was too  much noises for high interval time to fit
that curve.  In fact  errors on $\mathcal{R}_0$ are linear with $t$.   So we have smoothed
the curve  before fitting  by averaging  near points.  $\mathcal{R}_0$  for this  user was
fitted with a sum of two exponentials.

%%% \begin{eqnarray}
%%%   \lefteqn{e^{-0.197-8.12e-06t} + e^{-2.04-4.002e-04t}} \nonumber \\
%%%   & & \mbox{} + e^{-3.234-4.308e-03t}
%%% \end{eqnarray}. 

It seems that more than  the \emph{Login}/\emph{Logout} behavior there is also a
notion of user activity.  For example during the preparation of jobs or analysis
phase of the results an user does  not use the Grid and consequently the cluster
at all.  More  than the \emph{Login} and \emph{Logout}  state an \emph{Inactive}
state could be added to the model if needed.

\subsection{Other workloads comparison}

%% Particularité des jobs ne se trouve pas dans les autres modeles. 

 User number 3 is the most  active user from the NASA Ames iPSC/860 workload~\footnote{The
 workload log  from the NASA Ames iPSC/860  was graciously provided by  Bill Nitzberg. The
 workload  logs  from DAS2  were  graciously  provided by  Hui  Li,  David  Groep and  Lex
 Wolters. The  workload log from the SDSC  Blue Horizon was graciously  provided by Travis
 Earheart  and  Nancy   Wilkins-Diehr.   All  are  available  at   the  Parallel  Workload
 Archives~\url{http://www.cs.huji.ac.il/labs/parallel/workload/}\label{foot:pwa}}.
 Figure~\ref{fig:nasa:nasa}  shows  its  $\mathcal{R}_0(t)$  probability,  it  is  clearly
 hyper-exponential of  order 2, as other  users like number  22 and 23.  Other  users like
 number 12 and 15 are more classical Poissonian users.

 DAS2 Clusters (see  note \ref{foot:pwa}) used also PBS and  MAUI as their batch
 system and scheduler.   The main difference we have with them  is that they use
 Globus to co-allocate  nodes on different clusters.  We only  have bag of tasks
 applications which interacts together in a pipeline way by files stored on SEs.
 Their     fs0     144     CPUs     cluster    is     quite     similar     with
 ours. Figure~\ref{fig:nasa:das2}  shows the $\mathcal{R}_0(t)$  probability for
 their most active user and corresponding hyper-exponential fitting or order 2.

 SDSC Blue Horizon cluster (see note  \ref{foot:pwa}) have a total of 144 nodes.
 The $\mathcal{R}_0(t)$  distribution probability of their most  active user was
 fitted with a hyper-exponential of order 2 in figure~\ref{fig:nasa:sdsc}.
 
%% test.12.r0 ~ exp = Poisson process
%% test.14.r0 ~ droite => envoi regulier
%% test.24.r0 ~ hyper-exp

\section{Related works}
\label{related}

 Our  Grid environment  is very  particular  and different  from common  cluster
 environment as  parallelism involved requires no  interaction between processes
 and degree of parallelism is one for all jobs.

 To be  able to completely  simulate the  node usage we  need not only  the jobs
 submittal  process but also  the job  duration process.   Our runtime  model is
 similar  with  the Downey  model~\cite{downey99elusive}  for  runtime which  is
 composed of linear  pieces in logspace.  There is  a strong correlation between
 successive jobs  running time but  it seems unlikely  that a general  model for
 duration may be  made because it depends highly on algorithms  and data used by
 users.

 Most other models use  Poisson distribution for interarrival distribution.  But
 evidences,  like CV  be  much  higher than  one,  demonstrate that  exponential
 distributions does not fit  well the real data~\cite{feitelson01, jann97}.  The
 need  of a  detailed  model was  expressed  in~\cite{talby99b}.  With  constant
 parameters our model exhibits a hyper-exponential distribution for interarrival
 rate and justify  such a distribution choice.  One strong  benefit of our model
 is  that  it  is  general  and  could be  used  numerically  with  non-constant
 parameters at the expense of difficult fitting.
 
\section{Discussion}
\label{Discussion}

 What  could be  stated is  that job  maximum run  times provided  by  users are
 essentially inaccurate,  some authors are  even not using this  information for
 scheduling~\cite{darincosts}.  Maybe  a better concept is  the relative urgency
 of a job.   For example on a grid software managers  are people responsible for
 installing software  on cluster nodes  by sending installation  jobs.  Software
 manager jobs may  be regarded as more urgent than other  jobs type.  So sending
 jobs  with an  estimated runtime  could  be replaced  by sending  jobs with  an
 urgency parameter.  That urgency could be established in part as a site policy.
 Each site administrator  could define some users classes  for different kind of
 jobs and  software used  with different jobs  priorities.  For instance  a site
 hosted in some laboratory might wish to promote its scientific domain more than
 other domain, or some specific applications might need quality of services like
 real time interaction.

 Another idea for scheduling  is to have some sort of risk  assessment measured during the
 scheduler decision.  This risk assessment may be based on blocking probabilities obtained
 either from the logs or from some user behavior models.  For example, it could be wise to
 forbid that a group or  an user takes all the cluster at a given  time but instead to let
 some few percents of it open for short jobs or low CPU consuming jobs like monitoring.
 
 Information System  shows for a  site the number  of job currently  running and
 waiting. But it is not really the relevant metric in an on-line environment.  A
 better metric for a cluster is  the computing flow rate input and the computing
 flow rate capacity.  A cluster is  able to treat some amount of computation per
 unit of time.   So a cluster is contributing to the  Grid with some computation
 flow rate (in GigaFLOPS or TeraFLOPS).  As with classical queuing theory if the
 input rate is  higher than the capacity, the site is  overloaded and the global
 performances are low due to jobs  waiting to be processed. What happens is that
 the site  receive more jobs that  is is able to  treat in a given  time. So the
 queues begin to grow and jobs have  to wait more and more before being started,
 resulting in performance decay.   Similarly when the computation submitted rate
 is lower  than the site  capacity the site  is under-used.  Job  submittal have
 also to be  fairly distributed according to the site  capacity.  For example, a
 site  that  is twice  bigger  than  another site  have  to  receive twice  more
 computing request  than the  other site.   But there is  a problem  to globally
 enforce  this submittal  scheme on  all the  Grid.  This  is why  a  local site
 migration policy  may be better than  a central migration policy  done with the
 RB.

 To be more precise there are  two different kinds of cluster flow rate metrics,
 one is  the local flow  rate and the  other one is  the global flow  rate.  The
 local computing flow rate is the flow  rate that one job sees when reaching the
 site.  The global flow rate is the computing flow rate a group of jobs see when
 reaching the  site.  That  global flow  rate is also  the main  measurement for
 meta-scheduling between  sites.  These two metrics are  different, for instance
 we could have a  site with a lot of slow machines (low  local flow rate and big
 global flow rate) and another site with only few supercomputers (big local flow
 rate and low global flow rate).  But the most interesting metric for one job is
 the local  flow rate.   This means that  if each  job wants individually  to be
 processed at  the best  local flow rate  site, this  site will saturate  and be
 globally slow.

 As far as all  users and groups computation total flow rate is  less than the site global
 flow rate or site capacity, there is no real fairness issue because there is no strife to
 access the site  resources, there is enough for  all.  The problem comes when  the sum of
 all  computation flow  rate is  greater  than the  site capacity,  firstly this  globally
 reduces the site  performance, secondly the scheduler must take  decision to share fairly
 these resources.  The Grid is an ideal  tool that would allow to balance the load between
 sites by migrating jobs~\cite{darincosts}.  A site  that share their resources and is not
 saturated could  discharge another  heavily loaded  site.  Some kind  of local  site flow
 control could  maintain a bounded input  rate even with fluctuating  jobs submittal.  For
 instance fairness  between groups and  users could be  maintained by decreasing  the most
 demanding input rate and distributing it to other less saturated sites.

 Another  benefits is  that  applications  computing flow  rates  may be  partly
 expressed by users  in their job requirements.  Computing  flow rate takes into
 account both  the jobs sizes and  their time limits. Fairness  between users is
 then ensured if whatever may be flow values asked by each user, part granted to
 each  penalizes no  other one.  Computing flow  rate granted  by a  site  to an
 application may depend  on the applications degree of  parallelism, that is for
 the moment the number of jobs.  For  instance it may be more difficult to serve
 an application composed of only one job asking for a lot of computing flow rate
 than to serve  an application asking the same computing  flow rate but composed
 of many jobs.   Urgency is not totally measured by a  computing flow rate.  For
 example  a critical  medical application  which is  a matter  of life  or death
 arriving on  a full site has to  be treated in priority.  Allocating flow rates
 between users and groups has to be  right and to take under account priority or
 urgency issues.

 To use a site  wisely users have to bound their computational  flow rate and to
 negotiate  it with  site managers.   A computing  model has  to be  defined and
 published.  These remarks  are important in the case  of on-line computing like
 Grids  where meta-scheduling strategy  have to  take a  lot of  parameters into
 account.      General     on-line     load     balancing     and     scheduling
 algorithms~\cite{azar97line,   azar94line,  bar-noy00new,  lam02line}   may  be
 applied.  The problem of finding the  best suited scheduling policy is still an
 open problem.  A better understanding of  job running time is necessary to have
 a full model.

 The LCG middleware allows  users to send their jobs to different  nodes.  This is done by
 the way of a central element called  a Resource Broker, that collects user's requests and
 distributes  them  to computing  sites.   The  main purpose  is  to  match the  available
 resources and balance the load of  job submittal requests. Jobs are better localized near
 the data they need to use.

 We would  like to advise instead a  peer to peer~\cite{andrade-ourgrid} view  of the Grid
 over a centralized one.  In this view computing sites themselves work together with other
 computing  sites to  balance the  average workload.   Not relying  on  dependent services
 greatly improves  the reliability and  adaptability of the  whole systems.  That  kind of
 meta-scheduling have to  be globally distributed as stated by Dmitry  Zotkin and Peter J.
 Keleher~\cite{zotkin99joblength}:
 
 %% \begin{quote}
 \textit{In a distributed system like  Grid, the use of a central Grid
 scheduler}\footnote{like the Resource Broker used in LCG middleware}
 \textit{may result in a performance  bottleneck and lead to a failure
 of  the  whole  system.   It   is  therefore  appropriate  to  use  a
 decentralized scheduler architecture  and distributed algorithm.}  
 %%  \end{quote}  %% %%  \hfill{\small  (Dmitry  Zotkin  and Peter J. Keleher )}

 gLite~\cite{glite-design} is the next generation middleware for Grid computing.
 gLite will provide  lightweight middleware for Grid computing.   The gLite Grid
 services  follow   a  Service  Oriented  Architecture   which  will  facilitate
 interoperability among  Grid services.  Architecture details of  gLite could be
 viewed  in~\cite{glite-arch}.   The architecture  constituted  by  this set  of
 services is not bound to  specific implementations of the services and although
 the  services are expected  to work  together in  a concerted  way in  order to
 achieve the goals of the end-user  they can be deployed and used independently,
 allowing  their   exploitation  in  different  contexts.    The  gLite  service
 decomposition  has been largely  influenced by  the work  performed in  the LCG
 project.  Service implementations need to  be inter-operable in such a way that
 a client may talk to different independent implementations of the same service.
 This  can be  achieved in  developing  lightweight services  that only  require
 minimal  support from their  deployment environment  and defining  standard and
 extensible communication protocols between Grid services.
 
\hfill

 \section{Conclusion}
 \label{Conclusion}

 So far  we have analyzed the  workload of a Grid  enabled cluster and
 proposed an infinite Markov-based model that describes the process of
 jobs arrival.   Then a  numerical fitting has  been done  between the
 logs and the  model. We find a very similar  behavior compared to the
 logs, even bursts were observed during the simulation.

 \section*{Acknowledgments\markboth{Acknowledgments}{Acknowledgments}}
 The cluster at LPC  Clermont-Ferrand was funded by Conseil R\'egional
 d'Auvergne   within   the   framework   of  the   INSTRUIRE   project
 (\url{http://www.instruire.org})

\begin{small}

  \bibliographystyle{unsrt}
  \bibliography{workloadAnalysis.bib}

\begin{thebibliography}{10}

\bibitem{feitelson02workload}
Dror~G. Feitelson.
\newblock Workload modeling for performance evaluation.
\newblock In Maria~Carla Calzarossa and Salvatore Tucci, editors, {\em
  Performance Evaluation of Complex Systems: Techniques and Tools}, pages
  114--141. Springer-Verlag, Sep 2002.
\newblock Lect. Notes Comput. Sci. vol.~2459.

\bibitem{darincosts}
Darin England and Jon~B. Weissman.
\newblock Costs and benefits of load sharing in the computational grid.
\newblock In Dror~G. Feitelson and Larry Rudolph, editors, {\em Job Scheduling
  Strategies for Parallel Processing}. Springer-Verlag, 2004.

\bibitem{Garey79:intractability}
M.~Garey and D.S. Johnson.
\newblock {\em Computers and Intractability: A Guide to the Theory of
  {NP}-Completeness}.
\newblock Freeman, San Francisco, CA., 1979.

\bibitem{mertens04}
Stephan Mertens.
\newblock The easiest hard problem: Number partitioning.
\newblock In A.G. Percus, G.~Istrate, and C.~Moore, editors, {\em Computational
  Complexity and Statistical Physics}, New York, 2004. Oxford University Press.

\bibitem{school-parallel}
Dror~Feitelson School.
\newblock Parallel job scheduling --- a status report.
\newblock In Dror~G. Feitelson, Larry Rudolph, and Uwe Schwiegelshohn, editors,
  {\em Job Scheduling Strategies for Parallel Processing}, pages 1--16.
  Springer Verlag, 2004.

\bibitem{feitelson95d}
Dror~G. Feitelson and Larry Rudolph.
\newblock Parallel job scheduling: Issues and approaches.
\newblock In Dror~G. Feitelson and Larry Rudolph, editors, {\em Job Scheduling
  Strategies for Parallel Processing}, pages 1--18. Springer-Verlag, 1995.
\newblock Lect. Notes Comput. Sci. vol.~949.

\bibitem{jackson01}
David Jackson, Quinn Snell, and Mark Clement.
\newblock Core algorithms of the {Maui} scheduler.
\newblock In Dror~G. Feitelson and Larry Rudolph, editors, {\em Job Scheduling
  Strategies for Parallel Processing}, pages 87--102. Springer Verlag, 2001.
\newblock Lect. Notes Comput. Sci. vol.~2221.

\bibitem{linux-usenix}
Brett Bode, David~M. Halstead, Ricky Kendall, and Zhou Lei.
\newblock The {Portable Batch Scheduler} and the {Maui Scheduler on Linux
  Clusters}, {USENIX Association}.
\newblock {\em 4th Annual Linux Showcase Conference}, 2000.

\bibitem{geant4}
S.~Agostinelli et~al.
\newblock Geant 4 {(GEometry ANd Tracking)}: a {Simulation} toolkit.
\newblock {\em Nuclear Instruments and Methods in Physics Research}, pages
  250--303, 2003.

\bibitem{foster97globus}
Ian Foster and Carl Kesselman.
\newblock {Globus}: A metacomputing infrastructure toolkit.
\newblock {\em The International Journal of Supercomputer Applications and High
  Performance Computing}, 11(2):115--128, Summer 1997.

\bibitem{glite-arch}
EGEE~Design Team.
\newblock {EGEE} middleware architecture, {EGEE-DJRA1.1-476451-v1.0}, August
  2004.
\newblock Also available as~\url{https://edms.cern.ch/document/476451/1.0}.

\bibitem{zotkin99joblength}
Dmitry Zotkin and Peter~J. Keleher.
\newblock Job-length estimation and performance in backfilling schedulers.
\newblock In {\em {HPDC}}, 1999.

\bibitem{lcguserguide}
Antonio~Delgado Peris, Patricia~M\'endez Lorenzo, Flavia Donno, Andrea
  Sciab\`a, Simone Campana, and Roberto Santinelli.
\newblock {LCG User guide}, 2004.

\bibitem{dataGrid-wms}
G.~Avellino, S.~Beco, B.~Cantalupo, A.~Maraschini, F.~Pacini, M.~Sottilaro,
  A.~Terracina, D.~Colling, F.~Giacomini, E.~Ronchieri, A.~Gianelle, R.~Peluso,
  M.~Sgaravatto, A.~Guarise, R.~Piro, A.~Werbrouck, D.~Kou{\v{r}}il,
  A.~K{\v{r}}enek, L.~Matyska, M.~Mula{\v{c}}, J.~Posp{\'{i}}{\v{s}}il,
  M.~Ruda, Z.~Salvet, J.~Sitera, J.~{\v{S}}krabal, M.~Voc{\.{u}}, M.~Mezzadri,
  F.~Prelz, S.~Monforte, and M.~Pappalardo.
\newblock The datagrid workload management system: Challenges and results.
\newblock {\em Kluwer Academic Publishers}, 2004.

\bibitem{feitelson96c}
Dror~G. Feitelson and Larry Rudolph.
\newblock Toward convergence in job schedulers for parallel supercomputers.
\newblock In Dror~G. Feitelson and Larry Rudolph, editors, {\em Job Scheduling
  Strategies for Parallel Processing}, pages 1--26. Springer-Verlag, 1996.
\newblock Lect. Notes Comput. Sci. vol.~1162.

\bibitem{chiang02}
Su-Hui Chiang, Andrea Arpaci-Dusseau, and Mary~K. Vernon.
\newblock The impact of more accurate requested runtimes on production job
  scheduling performance.
\newblock In Dror~G. Feitelson, Larry Rudolph, and Uwe Schwiegelshohn, editors,
  {\em Job Scheduling Strategies for Parallel Processing}, pages 103--127.
  Springer Verlag, 2002.
\newblock Lect. Notes Comput. Sci. vol.~2537.

\bibitem{calzarossa93workload}
Maria Calzarossa and Giuseppe Serazzi.
\newblock Workload characterization: {A} survey.
\newblock {\em Proc. IEEE}, 81(8):1136--1150, 1993.

\bibitem{chapin99b}
Steve~J. Chapin, Walfredo Cirne, Dror~G. Feitelson, James~Patton Jones,
  Scott~T. Leutenegger, Uwe Schwiegelshohn, Warren Smith, and David Talby.
\newblock Benchmarks and standards for the evaluation of parallel job
  schedulers.
\newblock In Dror~G. Feitelson and Larry Rudolph, editors, {\em Job Scheduling
  Strategies for Parallel Processing}, pages 67--90. Springer-Verlag, 1999.
\newblock Lect. Notes Comput. Sci. vol.~1659.

\bibitem{cirnecompr-model}
Walfredo Cirne and Francine Berman.
\newblock A comprehensive model of the supercomputer workload, 2001.

\bibitem{downey99elusive}
Allen~B. Downey and Dror~G. Feitelson.
\newblock The elusive goal of workload characterization.
\newblock {\em Perf. Eval. Rev.}, 26(4):14--29, 1999.

\bibitem{feitelson95e}
Dror~G. Feitelson and Bill Nitzberg.
\newblock Job characteristics of a production parallel scientific workload on
  the {NASA Ames iPSC/860}.
\newblock In Dror~G. Feitelson and Larry Rudolph, editors, {\em Job Scheduling
  Strategies for Parallel Processing}, pages 337--360. Springer-Verlag, 1995.
\newblock Lect. Notes Comput. Sci. vol.~949.

\bibitem{paxson95wide}
Vern Paxson and Sally Floyd.
\newblock Wide area traffic: the failure of {Poisson} modeling.
\newblock {\em IEEE\slash ACM Transactions on Networking}, 3(3):226--244, 1995.

\bibitem{david-workload}
Hui Li, David Groep, and Lex Wolters.
\newblock Workload characteristics of a multi-cluster supercomputer.
\newblock In Dror~G. Feitelson, Larry Rudolph, and Uwe Schwiegelshohn, editors,
  {\em Job Scheduling Strategies for Parallel Processing}. Springer Verlag,
  2004.

\bibitem{kelsey98revised}
Richard Kelsey, William Clinger, and Jonathan~Rees (Editors).
\newblock Revised$^{5}$ report on the algorithmic language {Scheme}.
\newblock {\em ACM SIGPLAN Notices}, 33(9):26--76, 1998.

\bibitem{feitelson01}
Dror~G. Feitelson.
\newblock Metrics for parallel job scheduling and their convergence.
\newblock In Dror~G. Feitelson and Larry Rudolph, editors, {\em Job Scheduling
  Strategies for Parallel Processing}, pages 188--205. Springer Verlag, 2001.
\newblock Lect. Notes Comput. Sci. vol.~2221.

\bibitem{jann97}
Joefon Jann, Pratap Pattnaik, Hubertus Franke, Fang Wang, Joseph Skovira, and
  Joseph Riodan.
\newblock Modeling of workload in {MPP}s.
\newblock In Dror~G. Feitelson and Larry Rudolph, editors, {\em Job Scheduling
  Strategies for Parallel Processing}, pages 95--116. Springer Verlag, 1997.
\newblock Lect. Notes Comput. Sci. vol.~1291.

\bibitem{talby99b}
David Talby, Dror~G. Feitelson, and Adi Raveh.
\newblock Comparing logs and models of parallel workloads using the co-plot
  method.
\newblock In Dror~G. Feitelson and Larry Rudolph, editors, {\em Job Scheduling
  Strategies for Parallel Processing}, pages 43--66. Springer Verlag, 1999.
\newblock Lect. Notes Comput. Sci. vol.~1659.

\bibitem{azar97line}
Yossi Azar, Bala Kalyansasundaram, Serge~A. Plotkin, Kirk Pruhs, and Orli
  Waarts.
\newblock On-line load balancing of temporary tasks.
\newblock {\em J. Algorithms}, 22(1):93--110, 1997.

\bibitem{azar94line}
Yossi Azar, Andrei~Z. Broder, and Anna~R. Karlin.
\newblock On-line load balancing.
\newblock {\em Theoretical Computer Science}, 130(1):73--84, 1994.

\bibitem{bar-noy00new}
A.~Bar-Noy, A.~Freund, and J.~Naor.
\newblock New algorithms for related machines with temporary jobs.
\newblock In E.K. Burke, editor, {\em Journal of Scheduling}, pages 259--272.
  Springer-Verlag, 2000.

\bibitem{lam02line}
Tak-Wah Lam, Hing-Fung Ting, Kar-Keung To, and Wai-Ha Wong.
\newblock On-line load balancing of temporary tasks revisited.
\newblock {\em Theoretical Computer Science}, 270(1--2):325--340, 2002.

\bibitem{andrade-ourgrid}
Nazareno Andrade, Walfredo Cirne, Francisco Brasileiro, and Paulo Roisenberg.
\newblock {OurGrid}: An approach to easily assemble grids with equitable
  resource sharing.
\newblock In {\em Proceedings of the 9th Workshop on Job Scheduling Strategies
  for Parallel Processing}, June 2003.

\bibitem{glite-design}
{EGEE}~Design Team.
\newblock Design of the {EGEE} middleware grid services.
\newblock {\em EGEE JRA1}, 2004.
\newblock Also available as~\url{https://edms.cern.ch/document/487871/1.0}.

\end{thebibliography}

\end{small}

\end{document}